\documentclass[preprint,12pt]{elsarticle}

\usepackage{amssymb}

\usepackage{pdflscape}
\usepackage{siunitx}
\usepackage{tabularx}
\usepackage{booktabs}
\usepackage{graphicx} 
\usepackage{makecell}
\usepackage{xurl}
\usepackage{hyperref}
\graphicspath{ {./images/} }
\usepackage{subcaption}
\usepackage{textcomp}
\usepackage{float}
\usepackage{eurosym}

\journal{International Journal of Hydrogen Energy}

\begin{document}

\begin{frontmatter}



\title{Modeling Global Levelized Cost of Hydrogen Production Considering Country-Specific Investment Risks}


\author[inst1,inst2]{Stephan Kigle\corref{cor1}\fnref{fn1}}
\ead{skigle@ffe.de}

\author[inst1]{Tapio Schmidt-Achert\fnref{fn1}}
\ead{tschmidtachert@ffe.de}

\author[inst1]{Miguel Ángel Martínez Pérez}
\ead{mmartinez-perez@ffe.de}

\cortext[cor1]{Corresponding author}
\fntext[fn1]{Equal first authorship: both authors contributed equally to the paper.}

\affiliation[inst1]{organization={FfE Munich}, 
            addressline={Am Bluetenanger 71}, 
            city={Munich},
            postcode={80995}, 
            state={Bavaria},
            country={Germany}}

\affiliation[inst2]{organization={TUM}, 
            addressline={Arcisstraße 21}, 
            city={Munich},
            postcode={80333}, 
            state={Bavaria},
            country={Germany}}

\begin{abstract}

Hydrogen is central to the global energy transition when produced at low emissions. This paper introduces a renewable hydrogen production system model (HPSM) that optimizes a hybrid hydrogen production system (HPS) on a worldwide \qtyproduct{50x50}{\km} grid, considering country-specific interest rates. Besides the renewable energy's impact on the HPS design, we analyze the effect of country-specific interest rates on the levelized cost of hydrogen (LCOH) production. LCOH production ranges between 2.7~\geneuro/kg and 28.4~\geneuro/kg, with an average of 9.1~\geneuro/kg. Over one third (40.0\%) of all cells have an installed PV capacity share between 50\% and 70\%, and 76.4\% have a hybrid configuration. Hybrid HPSs can significantly reduce the LCOH production compared to non-hybrid designs, whereas country-specific interest rates lead to significant increases in the LCOH production. Hydrogen storage is deployed rather than battery storage to balance production and demand.
\end{abstract}

\begin{graphicalabstract}
\includegraphics[width=1\textwidth]{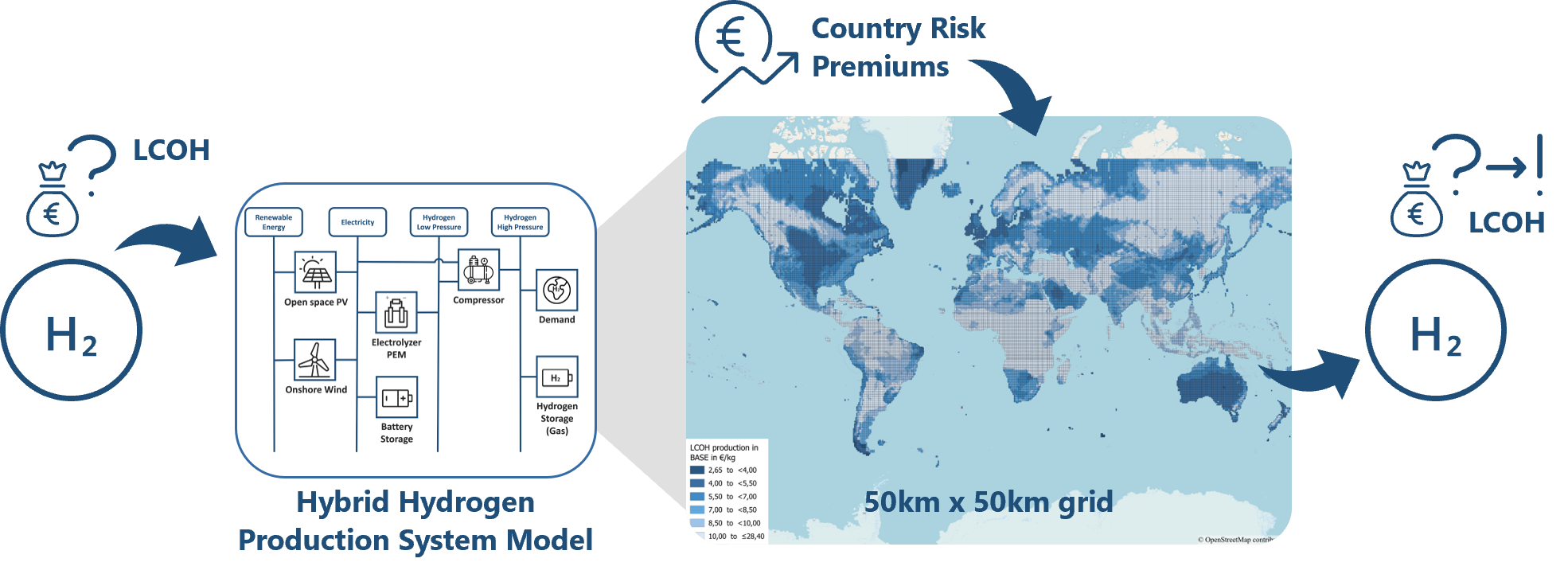}
\end{graphicalabstract}

\begin{highlights}
\item Worldwide levelized cost of hydrogen production on a \qtyproduct{50x50}{\km} grid resolution
\item Impact of country risk premiums on the levelized cost of hydrogen production
\item Cost-optimal hybrid hydrogen production system design per grid cell
\item All results are available on an open-data license
\end{highlights}


\end{frontmatter}


\section{Introduction}
\label{sec:Introduction}

As fossil fuels must be replaced to meet greenhouse gas (GHG) emission reduction targets, hydrogen is emerging as an essential energy carrier and feedstock to support global energy transition~\cite{Capurso.2022, Qazi.2022}. Common energy system studies show significant demand for hydrogen in different sectors and applications~\cite{Schmidt-Achert.2022,Wietschel.2021,Kigle.2022,IRENAPartI}. While providing the basis for clean power generation, electricity production from renewable energy sources such as wind onshore, wind offshore, or photovoltaic (PV) faces the problem of intermittency and transportability. Therefore, a low emissions, storable, and transportable energy carrier is needed to balance generation and consumption. Hydrogen can be such an energy carrier when produced from renewable electricity through electrolysis~\cite{IRENA.2022}.  Furthermore, hydrogen can be used as a feedstock and energy carrier to decarbonize several hard-to-abate applications that rely on fossil hydrocarbons. For this reason, all major economies are planning either to expand their domestic hydrogen production~\cite{IEAHydrogenProjects} or to import hydrogen from countries with favorable conditions for hydrogen production~\cite{IRENAPartI}, or both.

Unlike fossil fuels, only small amounts of direct GHG emissions and no CO\textsubscript{2} emissions are produced when hydrogen is combusted or re-converted into electricity via fuel cells. To reduce GHG emissions in total, it is all the more important that upstream emissions of hydrogen production, e.g., from electricity generation, are reduced to a minimum~\cite{reckrelevance, busch2023systematic}. For this reason, this paper focuses on producing green hydrogen via electrolysis using only electricity from renewable energy sources, particularly wind onshore and PV.

While the technical feasibility of producing hydrogen via electrolysis has been demonstrated many times already, there is great uncertainty regarding the price at which hydrogen will be available~\cite{Agora.2021}. Three components determine the total cost of hydrogen at a given point of consumption. Together, they form a global hydrogen model: the levelized cost of hydrogen (LCOH) production, the transportation cost, and the global demand for hydrogen. Combining all three elements with additional taxes and levies provides a comprehensive picture regarding the price at which hydrogen will be available at a given location. However, each component is very complex to analyze and requires different methods. This paper focuses on the first aspect, the LCOH production.

The LCOH production is defined as the ratio between a hydrogen production plant's discounted construction and operating costs and the expected hydrogen production volume during its lifetime. It is expressed in \geneuro/kg. The LCOH production is an evaluation parameter for comparing different hydrogen production technologies and localities~\cite{GovUKHydrogenCosts.2021}. The consistency of its calculation is essential for the comparability of the LCOH production between studies~\cite{AgoraUmlaut.2023}. The LCOH production is calculated as follows:

\begin{equation}
\label{equ:LCOH}
    LCOH = \frac{\sum_{i=1}^{i} AC_{i}}{PROD} \: in \: \frac{\euro{}}{kg_{H_{2}}}.
\end{equation}

$AC$ represents the total discounted annual cost of investments, including the operational cost for all system components $i$ (see Equation~\ref{equ:Annual Cost}) of the hydrogen production system (HPS). The annual costs are divided by the total amount of the annual hydrogen production $PROD$ of the HPS.

To calculate the LCOH production, the open source energy system model PyPSA~\cite{PyPSA} is used to configure an optimal HPS on a worldwide \qtyproduct{50x50}{\km} grid. The available components to model a cost-optimal HPS are the renewable energy technologies wind onshore and PV, polymer electrolyte membrane (PEM) electrolyzers, hydrogen storage units, hydrogen compressors, and stationary battery storage units (Section~\ref{sec:Methods}).\\

\begin{table}[]
    \footnotesize
    \centering
    \caption{Literature review of research articles calculating the LCOH production.}
    \label{tab:LCOH in Literature}
    \begin{tabular}{p{1cm} p{1cm} p{5.5cm} p{1.9cm} p{2.1cm}}
        \toprule
        \multicolumn{1}{m{1cm}}{Ref.} & \multicolumn{1}{m{1cm}}{Year} & \multicolumn{1}{m{5.5cm}}{Spatial Scope} & \multicolumn{1}{m{
        1.9cm}}{Electricity Source} & \multicolumn{1}{m{2.1cm}}{Method} \\
         \midrule

        \cite{Franzmann2023} & 2023 &  \qtyproduct 28 countries globally on country level & PV, WON & Optimization\\ 

        \cite{Janssen.2022} & 2021 &  30 European countries on country level & PV, WON, WOF & Analysis\\
        
        \cite{MDPI} & 2021 & point coordinates at 3 locations in Germany, Spain, and Morocco & PV, CSP & Optimization\\

        \cite{Lehmann.2021} & 2021 &  point coordinates in the state of Rio Grande do Norte, Brazil & PV & Analysis\\

        \cite{Khouya.2021} & 2021 &  point coordinates in Midelt, Morocco & CPV/T, CSP & Analysis\\

        \cite{Niaz.2021} & 2021 & point Coordinates in Incheon, Korea & PV+BESS & Analysis\\
         
        \cite{Multiobjective} & 2021 & point coordinates in Algiers, Algeria & WON & Optimization\\
        
        \cite{EWI} & 2020 & point coordinates in 94 countries & PV, WON, WOF & Optimization\\
                
        \cite{YatesJonathonRahman.2020} & 2020 & point coordinates in Australia, USA, Japan, Chile, and Spain & PV & Analysis\\

        \cite{Khouya.2020} & 2020 &  point coordinates in Tangier, Morocco & PV, WON & Analysis\\
        
        \cite{Gallardo.2020} & 2020 &  point coordinates in the Atacama Desert, Chile & PV, CSP+TES & Analysis\\

        \cite{Fasih} & 2020 &  \qtyproduct{0.45x0.45}{\degree} global grid & PV, WON & Optimization\\
        
        \cite{fromNorthafrica} & 2019 &  \qtyproduct{0.2x0.2}{\degree} grid in Algeria & WON, PV & Optimization\\
        
        \bottomrule
        \multicolumn{5}{p{13.6cm}}{\scriptsize Notes: PV = photovoltaic, CSP = concentrated solar power, WON = wind onshore, WOF = wind offshore, CPV/T = concentrated PV thermal system, TES =thermal energy storage, BESS = battery energy storage system}\\ 
    \end{tabular}
\end{table}

In literature, several studies are computing or analyzing the LCOH production for different geographical locations and spatial dimensions with various renewable energy technologies and methods. Table~\ref{tab:LCOH in Literature} compares these attributes. In addition to the articles listed in Table~\ref{tab:LCOH in Literature}, other non-peer-reviewed reports and tools that calculated the LCOH production have been published. In 2021 alone, three hydrogen potential atlases were published~\cite{PtXAtlas, H2AtlasAfrica.2022, AusH2} and a fourth is expected in 2024~\cite{HyPat}. Their focus is primarily on investigating the potential for hydrogen production, but the LCOH production is also calculated. The National Renewable Energy Laboratory (NREL)~\cite{HyDRA} published a similar tool for the US in 2009. While \cite{H2AtlasAfrica.2022, AusH2, HyDRA} analyze the LCOH production for different geographic locations and system configurations of renewable energy sources, \cite{Franzmann2023, Fasih, PtXAtlas, HyPat, IRENA.2022c} use optimization to determine a minimum cost system design for hydrogen generation. \cite{Fasih, PtXAtlas, HyPat} and \cite{IRENA.2022c} have a global scope, calculating the LCOH production on a generic grid~\cite{Fasih}, or focusing on specific locations~\cite{PtXAtlas}, or whole countries~\cite{HyPat, IRENA.2022c} for hydrogen production.

As this paper focuses on calculating an optimized HPS, Table~\ref{tab:Detail Literature} compares the research design of the articles labeled with the attribute 'Optimization' in Table~\ref{tab:LCOH in Literature}. The non-peer reviewed publications \cite{PtXAtlas}, \cite{IRENA.2022c}, \cite{AgoraAfry}, and \cite{AFRY.2022} are also included in the comparison, as their methodology offers additional insights.\\

As shown in Table~\ref{tab:Detail Literature}, approaches to calculating the LCOH production with optimization models vary widely. The most comprehensive approach is taken by \cite{PtXAtlas}. It investigates a combination of twelve different technology configurations at over 600 sites. Besides \cite{EWI}, only \cite{Fasih, PtXAtlas, IRENA.2022c}, and \cite{AFRY.2022} take a global approach. However, their analyses are not site-specific but provide the LCOH production for entire countries or regions, usually based on preselected locations. In contrast to \cite{EWI, Fasih, PtXAtlas, IRENA.2022c, AFRY.2022}, the other studies focus on individual regions~\cite{fromNorthafrica, AgoraAfry}, individual countries~\cite{Franzmann2023, Multiobjective}, or specific sites in different countries~\cite{MDPI}. \cite{Fasih} uses a similar global optimization approach as used in this paper, including a case study regarding the effect of different interest rates for two countries. In addition to the different spatial scopes, the analyzed studies differ regarding available renewable energy technologies and hydrogen production options.

Of the ten models in Table~\ref{tab:Detail Literature}, only two are nonlinear optimization models~\cite{MDPI, Multiobjective}. While \cite{MDPI} optimizes six, \cite{Multiobjective} optimizes four objective functions simultaneously to obtain a Pareto-optimal solution. In addition to minimizing the LCOH production, \cite{Multiobjective} also aims to minimize the curtailment of excess power and the unmet hydrogen demand. On the other hand, the avoided CO\textsubscript{2} emissions are maximized. All other models listed in Table~\ref{tab:Detail Literature} try minimizing LCOH production or ratios with different technology combinations and cost scenarios. Different model environments are used for this purpose. \cite{PtXAtlas}~uses the energy system model SCOPE, and \cite{MDPI} a global optimization tool implemented in MathWorks.

As most of the publications listed in Table~\ref{tab:Detail Literature}, we use linear optimization to model a HPS at minimal system costs. Therefore, we use the open source energy system model framework PyPSA~\cite{PyPSA}. The linear approach is favored as it is computationally less expensive, and the analysis focuses on the LCOH production. However, this paper differs significantly from the publications listed in Table~\ref{tab:Detail Literature} by its high spatial resolution and the global coverage of hydrogen production sites combined with the possibility of a hybrid HPS design and the consideration of specific country risk premiums (CRPs). To achieve this goal, PyPSA is combined with weather data from the Modern-Era Retrospective Analysis for Research and Application, Version 2 (MERRA-2) dataset from the NASA Global Modeling and Assimilation Office~\cite{NasaMerra} and CRP data from \cite{CRP.2022}. Thus, we can compute LCOH production on a \qtyproduct{50x50}{\km} grid worldwide. 

\begin{landscape}
\begin{table}[]
    \footnotesize
    \centering
    \caption{Research design in literature using optimization models to calculate the LCOH production.}
    \label{tab:Detail Literature}
    \begin{tabular}{p{0.7cm} p{1.5cm} p{2.8cm} p{1.7cm} p{0.7cm} p{0.7cm} p{0.7cm} p{0.7cm} p{1.5cm} p{3.6cm}}
        \toprule
        \multicolumn{1}{m{0.7cm}}{Ref.} & \multicolumn{1}{m{1.5cm}}{Product} & \multicolumn{1}{m{2.8cm}}{Spatial Scope} & \multicolumn{1}{m{1.7cm}}{ES} & \multicolumn{1}{m{0.7cm}}{HS} & \multicolumn{1}{m{0.7cm}}{GC} & \multicolumn{1}{m{0.7cm}}{ES} & \multicolumn{1}{m{0.7cm}}{CSC} & \multicolumn{1}{m{1.5cm}}{OM} & \multicolumn{1}{m{3.6cm}}{Weather Data}\\
         \midrule
        \cite{Franzmann2023} & H\textsubscript{2} & global, 28 countries & PV, WON & \checkmark & X & \checkmark &  \checkmark & linear & MERRA-2 and GWA (Wind)\\
        \hline
        \cite{MDPI} & H\textsubscript{2} & point coordinates (3) & PV, CSP & X & X & X & X & non-linear & Meteonorm\\
         \hline
        \cite{Multiobjective} & H\textsubscript{2} & point coordinates (1) & WON & X & \checkmark & \checkmark & X & non-linear & CDER Meteorological station in Algeria\\
         \hline
        \cite{EWI} & H\textsubscript{2} & point coordinates (94) & PV, WON, WOF & \checkmark & X & X & \checkmark & linear & SRB Release 3 (PV), MERRA-2 and GWA (Wind)\\ 
         \hline
        \cite{Fasih} & H\textsubscript{2} & global & PV, WON & \checkmark & X & \checkmark &  \checkmark & linear & SSE (NASA)\\
         \hline
        \cite{fromNorthafrica} & H\textsubscript{2} & \qtyproduct{0.2x0.2}{\degree}, Algeria & PV, WON & \checkmark & X & \checkmark & X & linear & CM-SAF SARAH (PV), MERRA-2 (Wind)\\
        \hline
        \cite{PtXAtlas} & H\textsubscript{2}, SNG, FT, Methanol & point coordinates (global) & PV, WON & \checkmark & X & \checkmark & X & linear & ERA5\\
         \hline
        \cite{IRENA.2022c} & H\textsubscript{2} & global & PV, WON, WOF & \checkmark & X & X &  \checkmark & linear & ERA5\\ 
         \hline
        \cite{AgoraAfry} & H\textsubscript{2} & hexagrid (50.000 \si{\square\km}) Europe and North Africa & PV, WON, WOF & \checkmark & X & X & X & linear & Solargis (PV), MERRA-2 and GWA (Wind)\\
         \hline
        \cite{AFRY.2022} & H\textsubscript{2} & global, 31 regions & PV, WON & X & X & X & X & linear & -\\
        \bottomrule
        \multicolumn{10}{p{17.2cm}}{\scriptsize Notes: ES = Electricity Source, HS = Hybrid System, GC = Grid Connection, ES = Electrical Storage, CSC = Country (or regional) Specific Costs, OM = Optimization Method}\\ 
    \end{tabular}
\end{table}
\end{landscape}

\section{Methods}
\label{sec:Methods}

To calculate the LCOH production on a global \qtyproduct{50x50}{\km} grid, an optimized HPS is modeled with PyPSA~\cite{PyPSA} for each grid cell. Since renewable electricity generation builds the basis of green hydrogen production, Section~\ref{subsec:Input Data and Spacial Resolution} describes the weather data and the resulting electricity production time series used as input for the hydrogen production system model (HPSM). We then specify the model components' techno-economic parameters and the implementation of the HPS in the PyPSA modeling framework in Section~\ref{subsec:Hydrogen Production System Model}. In Section~\ref{subsec:Country Risk Premiums}, we discuss the influence of CRPs on the LCOH production. Finally, Section~\ref{subsec:Computational Performance} explains the computational requirements.

\subsection{Weather Data, Spatial and Temporal Resolution}
\label{subsec:Input Data and Spacial Resolution}

\begin{figure}[h]
    \centering
    \includegraphics[width=0.8\textwidth]{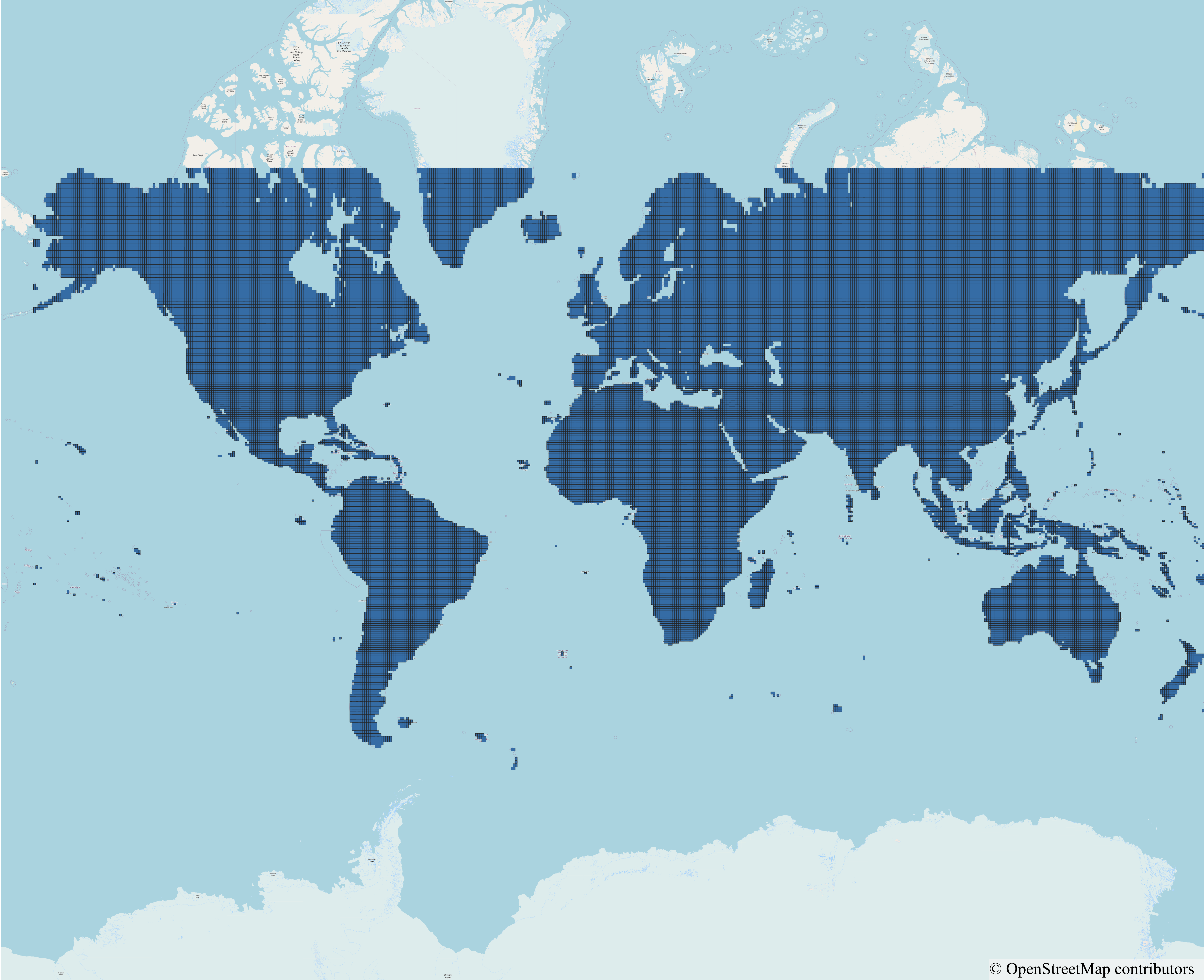}
    \caption{World map with MERRA-2 cells included in the HPSM.}
    \label{fig:Spatial Resolution}
\end{figure}

Renewable electricity generation highly depends on local weather conditions, such as solar radiation and wind speed. Therefore, the availability of reliable weather data limits the model's resolution. To model an electrolysis-based HPS, we use the MERRA-2 dataset from the NASA Global Modeling and Assimilation Office~\cite{NasaMerra}. The dataset provides meteorological data for a global grid with 207.936 cells. Each cell has a size of \qtyproduct{0.625x0.5}{\degree} (approx. \qtyproduct{50x50}{\km}). The meteorological data is available for each year since 1980. For the calculations in this paper, the weather year 2012 is used. Cells located offshore, in Antarctica, or at the North Pole are excluded from the analysis. There are two reasons to do so: the computational effort is reduced since only about one-quarter (51,677 cells) of all MERRA-2 cells are located inland; Antarctica and the North Pole, geographic regions situated far from populated areas with extreme weather conditions, are unlikely to become hydrogen production sites. Additionally, this study focuses on the LCOH production and optimal HPS design rather than production potentials (see Section~\ref{sec:Discussion}). Figure~\ref{fig:Spatial Resolution} shows the cells for which an optimized HPS is calculated.  

For the HPSM described in Section~\ref{subsec:Hydrogen Production System Model}, MERRA-2 data is used for each cell, firstly to assign the geographical location of each cell to a single country and secondly to account for meteorological variables such as solar radiation and hourly wind speed. The time series for the actual electricity produced by the renewable energy technologies wind onshore and optimally tilted PV are calculated as in \cite{WindData_MA}. Therefore, the HPSM has a spatial resolution of \qtyproduct{50x50}{\km} and a temporal resolution of one hour.

\subsection{Hydrogen Production System Model}
\label{subsec:Hydrogen Production System Model}

\begin{figure}[h]
    \centering
    \includegraphics[width=0.7\textwidth]{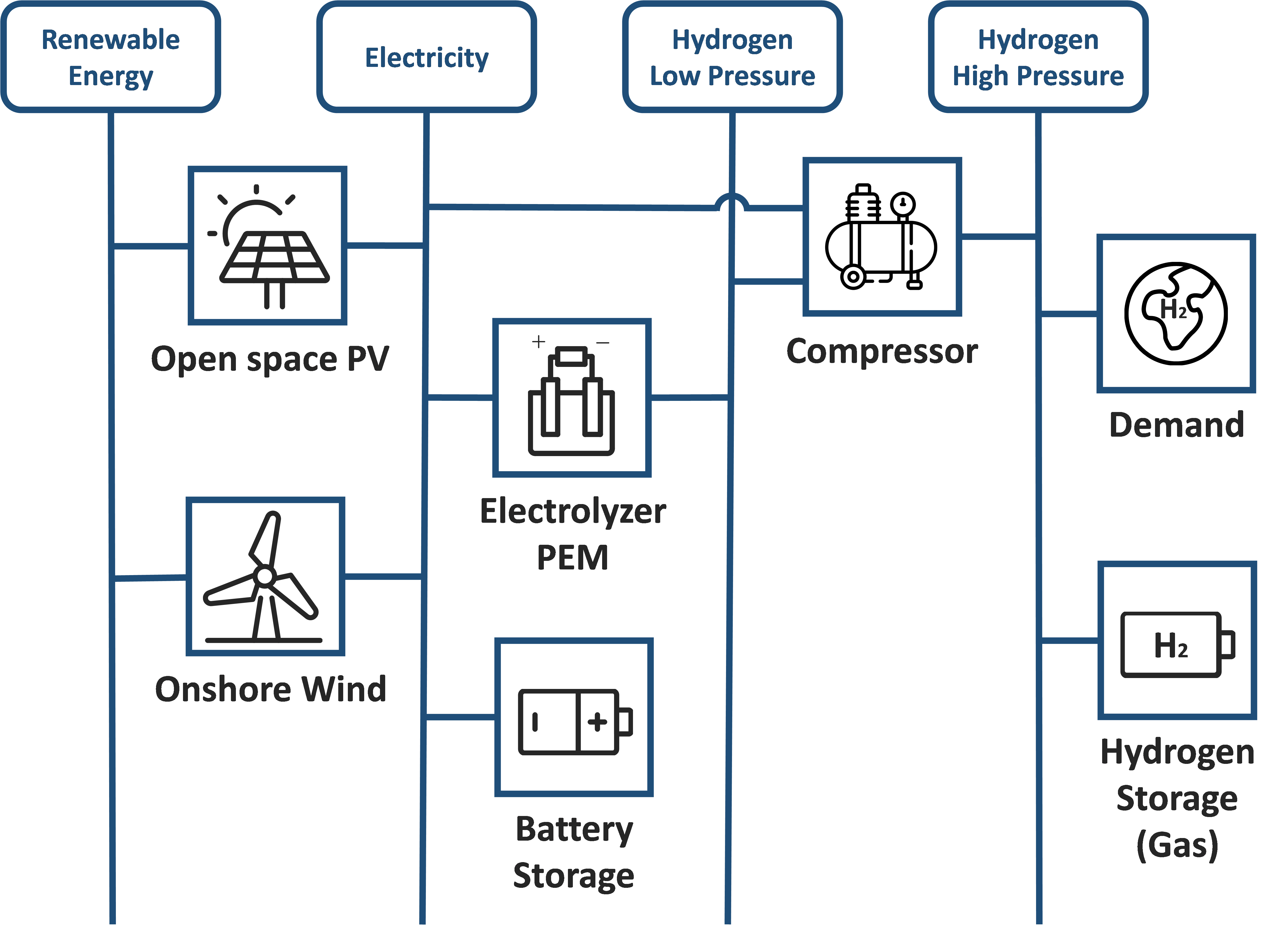}
    \caption{Hydrogen Production systems' model components.}
    \label{fig:Model Components}
\end{figure}

To meet the  hydrogen demand of 1~kg~H\textsubscript{2}/h in each cell, the model can combine the technologies shown in Figure~\ref{fig:Model Components} to minimize the total system's cost. The boundary condition imposed on the system is the requirement to produce 1~kg~H\textsubscript{2}/h in each cell constantly. While \cite{EWI, AFRY.2022} use a country or regional-specific hydrogen demand constraint, \cite{fromNorthafrica} arbitrarily sets the demand to 1~GWh/a. \cite{Multiobjective} also uses 1~kg~H\textsubscript{2}/h as the boundary condition for hydrogen production. \cite{Fasih} uses a constant small hourly demand. Since the LCOH production is calculated as a function of the amount of hydrogen produced (see Equation~\ref{equ:LCOH}), we suppose that the defined absolute demand only affects the LCOH production if the potential area of any required technology is exceeded. The extent to which the demand constraints may affect the LCOH production is discussed in Section~\ref{sec:Discussion}.   

The following technologies are available to the HPSM to build a cost-optimal HPS: ground-mounted PV,  onshore wind turbines (strong and weak wind turbines), PEM electrolyzers, hydrogen storage units, hydrogen compressors, and stationary battery storage units. The PEM electrolyzers are assumed to operate between 0\% and 100\% of rated power. Ramping up to rated power from standby condition is supposed to be possible within a few seconds~\cite{Milanzi.2018}. As the temporal resolution of the model is one hour, no additional restrictions for the electrolyzer operation must be implemented. The hydrogen storage system is a tubular storage system that can buffer hydrogen production and demand in the short term. The stationary battery storage units are modeled with an energy-to-power ratio of E/P = 4 and no discharge losses~\cite{McLaren.2022}.  All storage levels (hydrogen and battery storages) must equal the initial state of charge at the end of the optimization period of 8784~h. The compressor elevates the pressure of the hydrogen to 100 bar.

Additional components such as offshore wind turbines, desalination plants, heaters, or heat pumps for heat supply and the electrical transmission grid are not included in the model to minimize the necessary computational effort. Hydrogen production is assumed to occur off-grid without any connection to the local electrical grid. The techno-economic parameters for each technology used in the model can be found in Table~\ref{tab:Technoeconomic Parameters}.

\begin{table}[]
    \footnotesize
    \centering
    \caption{Techno-economic parameters for the HPSM's components wind onshore \cite{Guminski.2021} and own cost calculations, PV \cite{Guminski.2021}, H\textsubscript{2} Storage \cite{Reu.2019, Heuser.2021}, Battery Storage \cite{Guminski.2021, Fattler.2019}, PEM electrolyzer \cite{Fattler.2019, Smolinka.2018} and compressor \cite{Compressor.2015}. All prices are adjusted to represent real prices in 2020.}
    \label{tab:Technoeconomic Parameters}
    \begin{tabular}{p{2.8cm} p{1.7cm} p{1.7cm} p{1.9cm} p{1.3cm} p{1.3cm}}
        \toprule
         & \multicolumn{3}{c}{Economic Parameters} & \multicolumn{2}{c}{Technical Parameters}\\
            \cmidrule{2-6}
        Technology & CAPEX\textsubscript{1}\newline ($\euro{}/MW$) & CAPEX\textsubscript{2}\newline ($\euro{}/MWh$) & OPEX (\% of CAPEX\textsubscript{1}) & Lifetime ($a$) & Efficiency (\%) \\
            \midrule
        PV & 685 456 & - & 2.5\% & 25 & - \\
            \hline
        Weak Wind\newline Turbine & 2 034 400 & - & 1.4\% & 25 & - \\
            \hline
        Strong Wind\newline Turbine & 740 000 & - & 3.9\% & 25 & - \\
            \hline
        H\textsubscript{2} Storage & 1 518 & 15 179 & 2.0\% & 30 & 97.5\% \\
            \hline
        H\textsubscript{2} Compressor & 4 700 717 & - & 4.0\% & 30 & 97.5\% \\
            \hline
        Battery Storage & 530 410 & 138 229 & 5.8\% & 10 & 95.0\% \\
            \hline
        PEM Electrolyzer & 1 495 067 & - & 2.0\% & 20 & 58.0\% \\
        \bottomrule
        
    \end{tabular}
\end{table}

Since wind turbine power generation is highly dependent on local wind conditions, two different turbine types are included in the model: a weak wind turbine (Enercon E-115) and a strong wind turbine (Enercon E-82 E3). The tower heights of the used weak wind turbine is higher, and their power density is lower than that of strong wind turbines. Electricity from weak wind turbines is more constant at most locations. The disadvantage comes with higher investment costs and less power generation at higher wind speeds or gusts. In addition, not all sites are suitable for installing weak wind turbines, as excessively high wind speeds prevent safe operation. Therefore, we exclude the possibility of installing weak wind turbines in cells where the average wind speed and the reference wind speed are above the limits of wind class I according to the IEC 61400-1:2019 standard. To calculate a cost-optimal HPS, both types of turbines can be installed in cells where boundaries are not exceeded. 

The total annual discounted cost for each technology is used since the optimization is performed for 8784 time steps, i.e., one year. The period of one year is chosen to take into account the seasonality of electricity generation from renewable energy sources. The annual cost $AC$ for any technology $i$ is calculated as:

\begin{equation}
\label{equ:Annual Cost}
    AC_{i} = \frac{CAPEX_{i}}{AF} + OPEX_{i},
\end{equation}

where $AF$ is the annuity factor:

\begin{equation}
\label{equ:Annuity Factor} 
    AF = \frac{(1+r)^{n}-1}{r*(1+r)^{n}}.
\end{equation}

$n$ stands for the lifetime of the corresponding technology $i$. $r$ is the weighted average cost of capital (WACC) comprising two components. A baseline with $WACC_{const} = 3,5\%$ and a specific CRP for each country. Therefore, $r = WACC_{const} + CRP$. The nominal power is the optimization variable for each component; only for the hydrogen storage, the nominal capacity is independently optimized, too.

\subsection{Country Risk Premiums}
\label{subsec:Country Risk Premiums}

CRPs, or country-specific market risk premiums, account for the additional return on investment investors demand when investing in countries with higher risk. The CRP includes the risk of the political and economic instability of any country and can be estimated using average historical differential returns in comparison to "risk-free" rates of returns~\cite{Fernandez.2007}, the country's default risk reflected in bond default spreads~\cite{Damodaran.2022} or collected data from surveys of experts and companies in finance and economics~\cite{Fernandez.2022}. Despite the CRP's importance in evaluating corporate investments, no method has yet proven consistent enough to measure the total risk~\cite{Taylor.2021}.

We use CRPs derived from bond default spreads as calculated in \cite{CRP.2022} in our proposed HPSM to investigate their effect on the LCOH production. This method considers the bond default spreads, the difference between the interest rate of a country-specific treasury bond, and a reference "risk-free" treasury bond, both issued for the same time to maturity (ten years) and currency (in dollars). Additionally, a volatility factor incorporates the investment's short-term risk using the equity markets' standard deviation. This method estimates the CRP using current market indicators and sovereign ratings for those countries lacking market information~\cite{Damodaran.2022}.

\begin{equation}
\label{equ:Country Risk Premiums Bond default spread method}
    CRP = {Bond\;Default\;Spread} * {Volatility Factor},
\end{equation}

Considering CRPs in the corresponding models can lead to a more realistic assessment regarding the cost-effectiveness of hydrogen production in a specific country. CRPs for different countries range between 0\% and 19.18\% in 2020~\cite{CRP.2022}. For a few missing countries, the CRPs of neighboring countries were used.

Until now, CRPs and country-specific discount rates have not been globally and systematically used when calculating the LCOH production even though discount rates have a significant impact on the LCOH production~\cite{AgoraUmlaut.2023,Fasih}. Of the publications listed in Table~\ref{tab:Detail Literature}, only \cite{IRENA.2022c} considers CRPs, alternatively \cite{EWI, Fasih} consider regional CAPEX costs and a hypothetical WACC sensitivity analysis respectively.

\subsection{Computational Performance}
\label{subsec:Computational Performance}
The electrolysis-based HPS is implemented in the open-source Python environment PyPSA~\cite{PyPSA}, used for simulating and optimizing modern energy systems. PyPSA uses an optimization module based on linopy~\cite{Hofmann2023} as a substitute for pyomo~\cite{hart2011pyomo, bynum2021pyomo} to formulate each optimization problem. The solver Gurobi~\cite{gurobi} is used for its faster performance than free solvers. Despite the performance improvement due to the solver Gurobi~\cite{gurobi} reaching optimization solutions within 40~s for one cell, the computational time would add up to 24 days due to the number of cells (51,677), as each cell forms a separate optimization problem. However, since the optimization problems are not dependent on each other parallelization can reduce the computational time. Results were computed on a server with two CPUs, 46 virtual cores, and 469 GB RAM, using a maximum of 70\% or 32 virtual cores of the server, reducing the computational time for one scenario to about 36 h, as two cores are used per optimization.

\section{Results}
\label{sec:Results}

With the HPSM presented in Section~\ref{sec:Methods}, we calculate four scenarios with different model configurations to investigate the LCOH production. The four scenarios are

\begin{itemize}
    \item \textbf{cWACC:} optimizes the HPS with constant WACC of 3.5\% for all countries 
    \item \textbf{BASE:} same as cWACC but considers CRPs for each country 
    \item \textbf{PVonly:} same as BASE with the restriction that only PV is available as an electricity production technology 
    \item \textbf{WINDonly:} same as BASE with the restriction that only wind onshore is available as an electricity production technology 
\end{itemize}

Through the four scenarios, it is possible to investigate the impact of different parameters on the LCOH production. While cWACC allows us to compare the LCOH production of each cell based on physical conditions, BASE gives a more realistic picture regarding the LCOH production as country-specific risk premiums are reflected in the model results. Comparing BASE with PVonly and WINDonly allows us to determine the effects of hybrid configurations on the LCOH production. In the following, we will discuss and compare the results of the four scenarios.

\subsection{Scenario cWACC}

Figure~\ref{fig:cWACC and BASEa} shows the LCOH production in the scenario cWACC. The LCOH production ranges from 2.6~\geneuro/kg (Kerguelen Islands, France) to 25.7~\geneuro/kg (Canada), while the mean LCOH production is 7.2~\geneuro/kg and the median is 6.9~\geneuro/kg. The differences in the HPS design cause different LCOH production. Regions with low LCOH production are, for example, South America (Chile, Argentina), parts of the USA and Canada, Australia, or the Middle East. However, due to the high spatial resolution, it becomes clear that statements regarding the LCOH production in specific regions or countries can only be made with difficulty. The LCOH production can vary strongly within several hundred kilometers, i.e., in North America. This results from the varying meteorological conditions, as seen in Figure~\ref{fig:FLH cWACC and BASEa} and \ref{fig:FLH cWACC and BASEb}. PV and wind onshore full load hours (FLH) show a strong negative correlation with the LCOH production at a significance level of 1\%. The correlation coefficient of wind onshore is -0.71. In contrast, the correlation coefficient of PV is -0.38, meaning that the correlation of wind onshore and LCOH production is higher than for PV and the LCOH production. Our linear regression results show a negative relation on a 1\% significance level for both wind onshore and PV FLH and LCOH production. Hence, LCOH production is lowest when good PV, wind onshore, or both resources are available in one cell. This is why, in 76.4\% of the calculated cells, hybrid renewable electricity production systems are chosen for the cost-optimal HPS design. Only 21.0\% of the calculated cells use PV, and 2.6\% use wind as the single renewable energy source.

\subsection{Scenario BASE}

\begin{figure}
  \centering
  \begin{subfigure}{0.49\textwidth}
    \includegraphics[width=\linewidth]{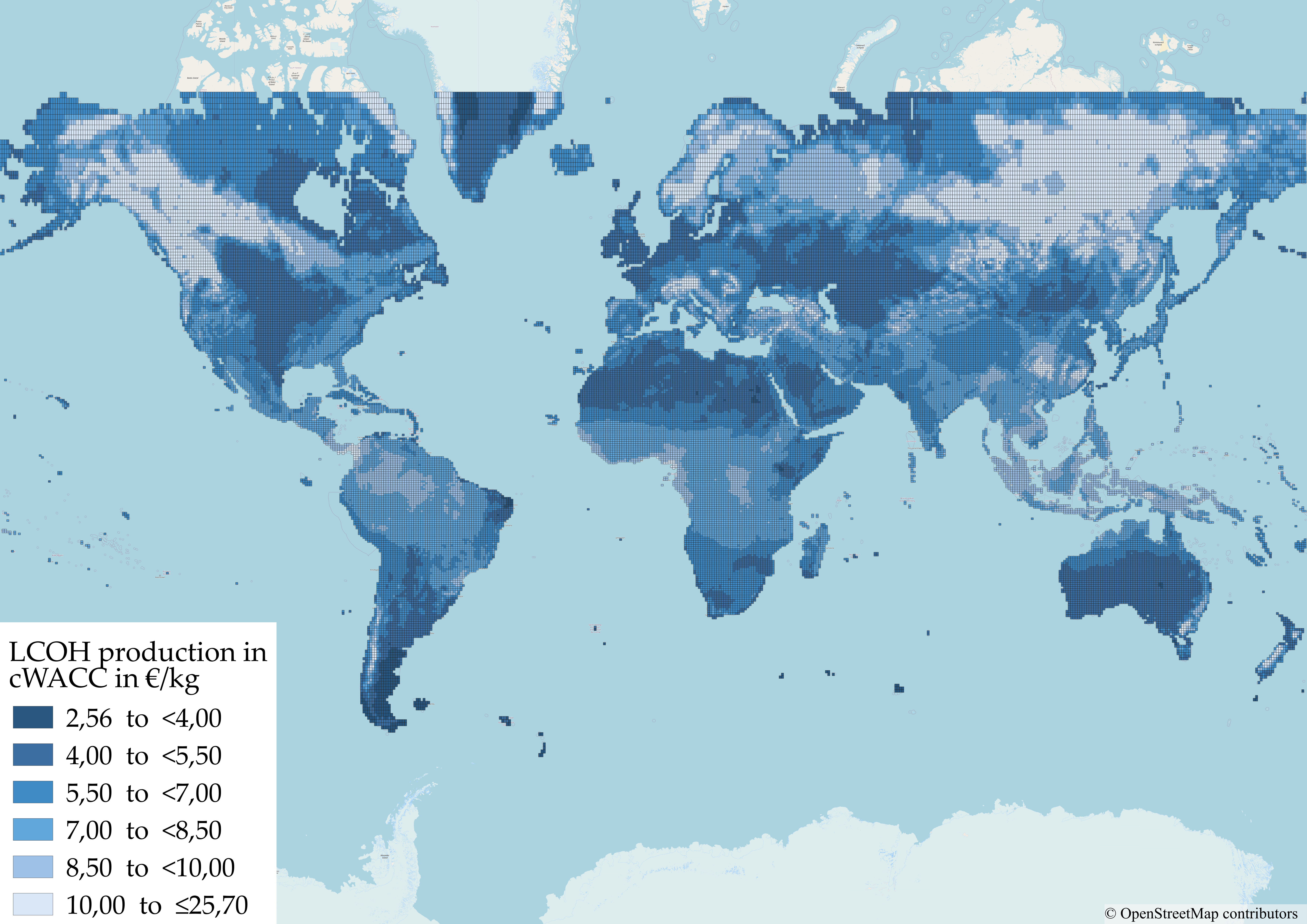} 
    \caption{Scenario cWACC} \label{fig:cWACC and BASEa}
  \end{subfigure}
  \hspace*{\fill}   
  \begin{subfigure}{0.49\textwidth}
    \includegraphics[width=\linewidth]{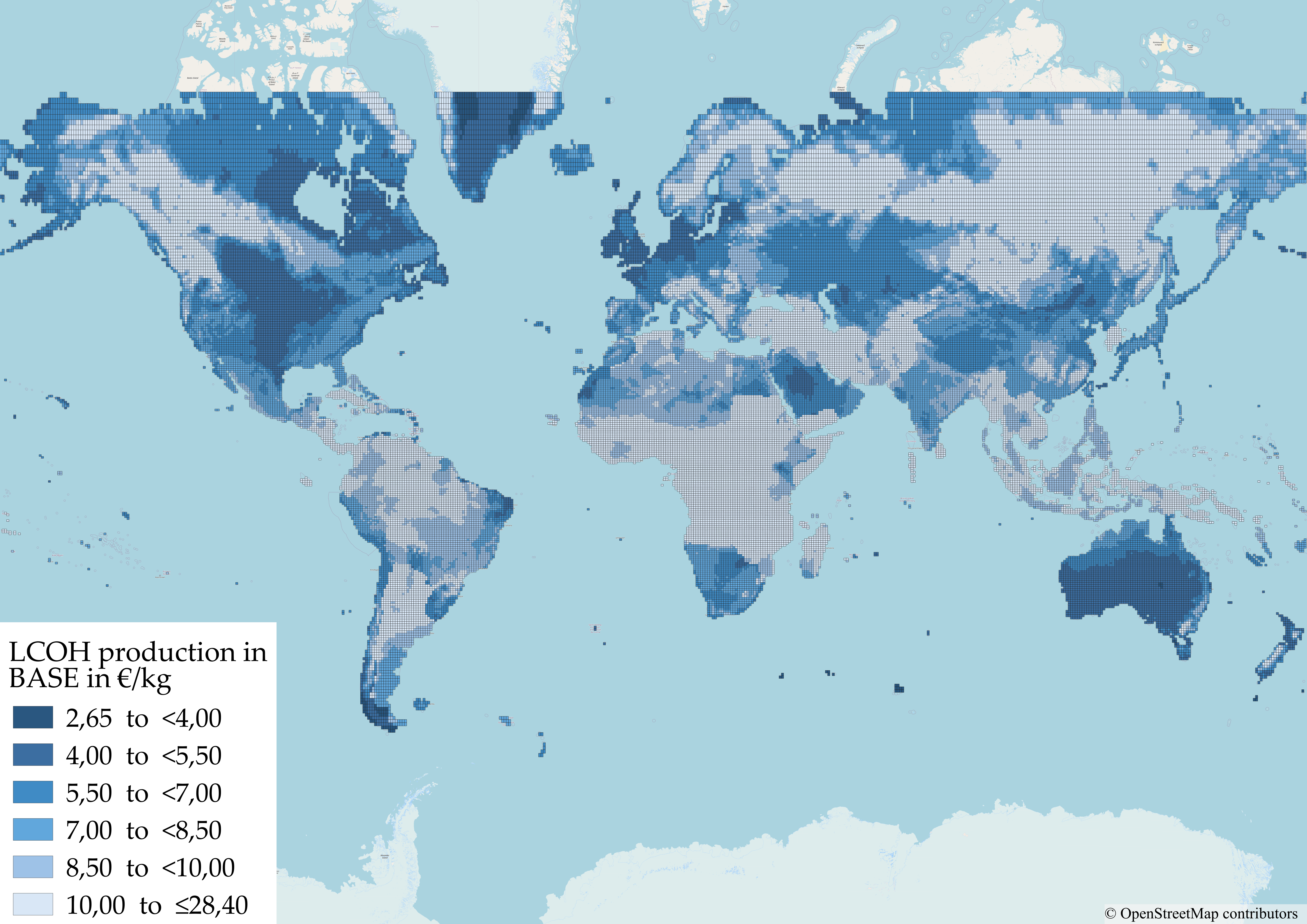} 
    \caption{Scenario BASE} \label{fig:cWACC and BASEb}
  \end{subfigure}

\caption{Global LCOH production on a \qtyproduct{50x50}{\km} grid for the scenarios cWACC and BASE.}
\label{fig:cWACC and BASE}
\end{figure}

\begin{figure}[h]
    \centering
    \begin{subfigure}[t]{0.49\textwidth}
    \includegraphics[width=\linewidth]{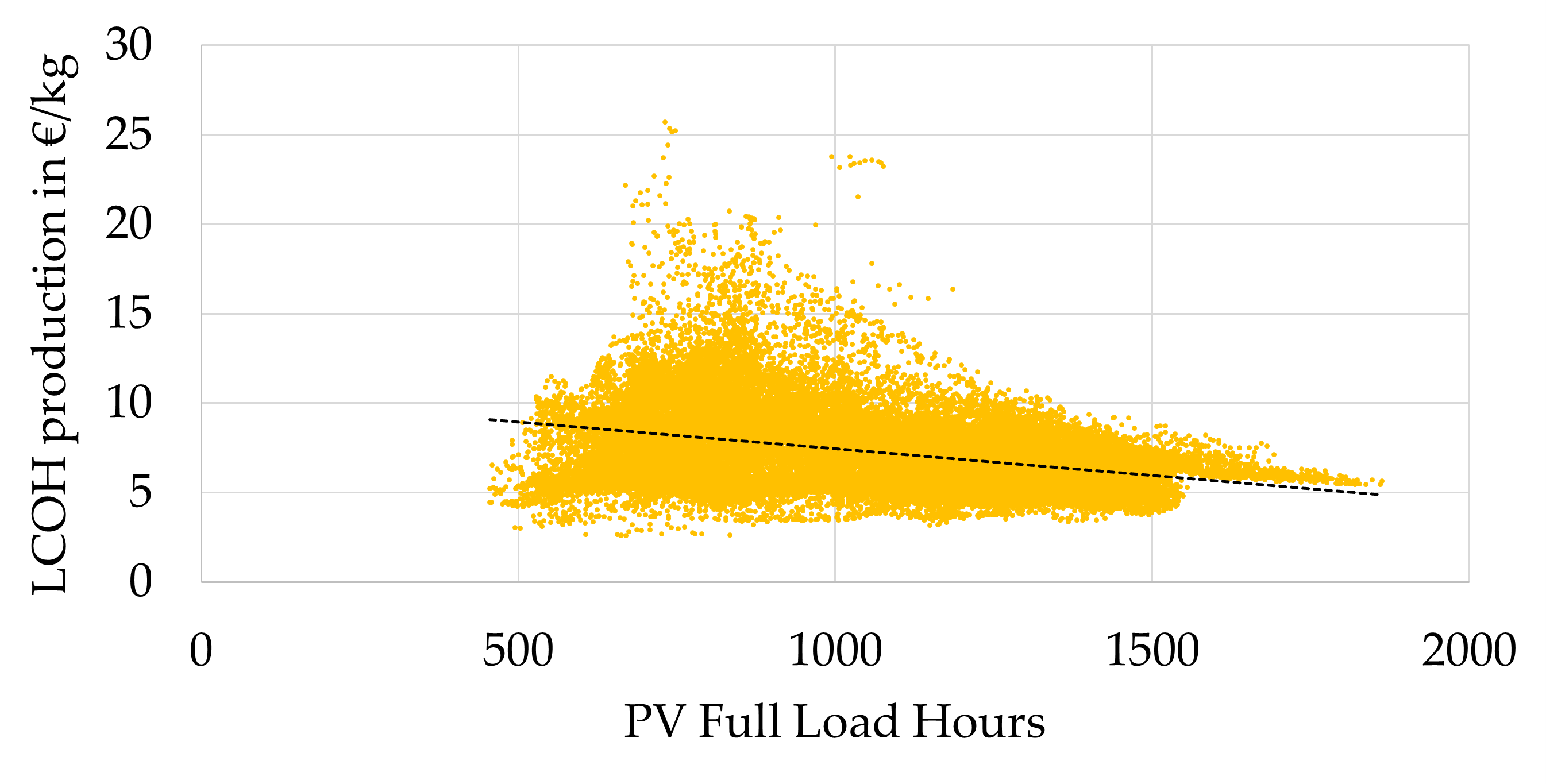} 
    \caption{FLH vs. LCOH production for PV in the scenario cWACC} \label{fig:FLH cWACC and BASEa}
  \end{subfigure}
  \hspace*{\fill}   
  \begin{subfigure}[t]{0.49\textwidth}
    \includegraphics[width=\linewidth]{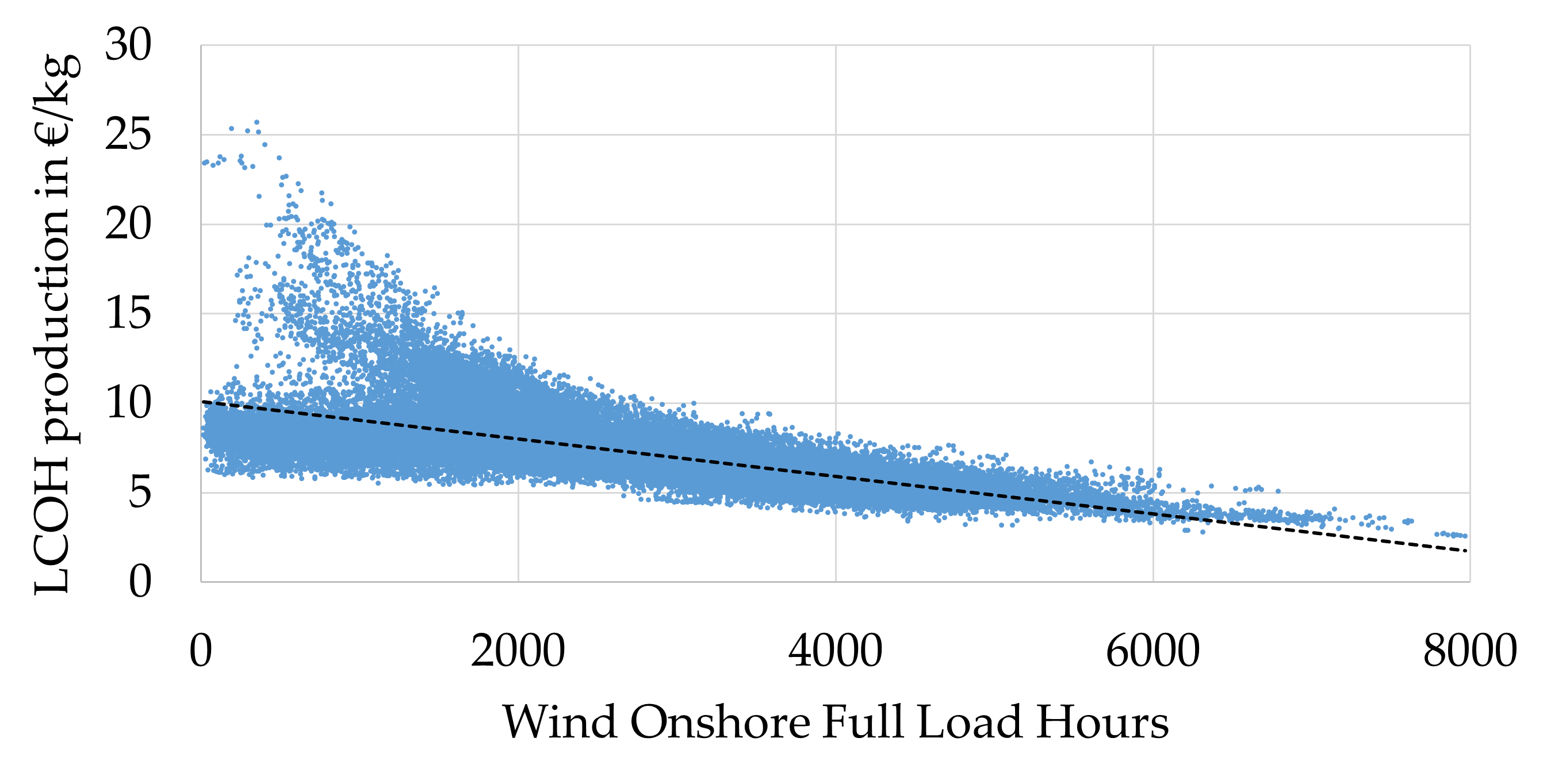} 
    \caption{FLH vs. LCOH production for wind onshore in the scenario cWACC} \label{fig:FLH cWACC and BASEb}
  \end{subfigure}
  \begin{subfigure}[b]{0.49\textwidth}
    \includegraphics[width=\linewidth]{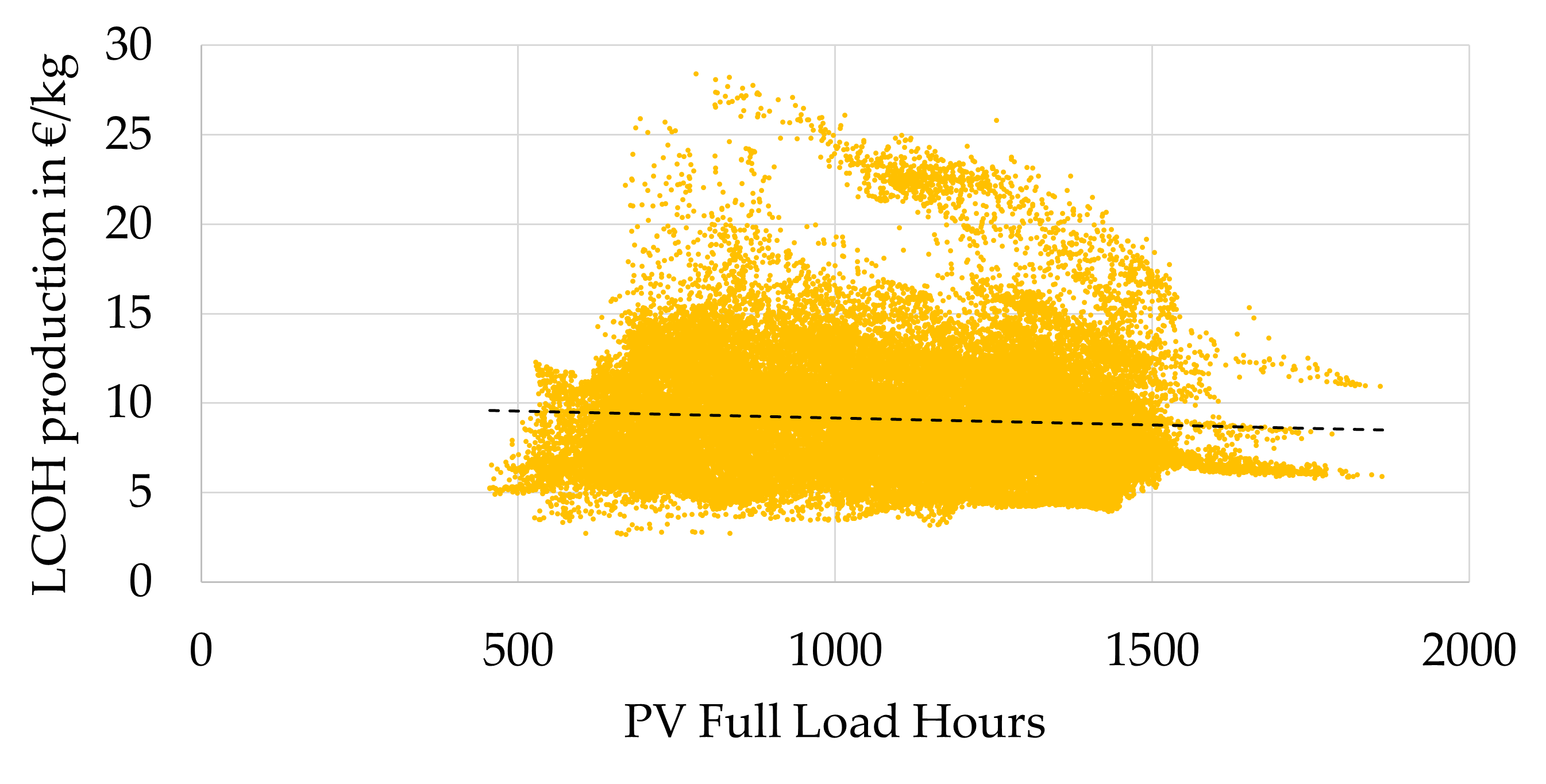} 
    \caption{FLH vs. LCOH production for PV in the scenario BASE} \label{fig:FLH cWACC and BASEc}
  \end{subfigure}
  \hspace*{\fill}   
  \begin{subfigure}[b]{0.49\textwidth}
    \includegraphics[width=\linewidth]{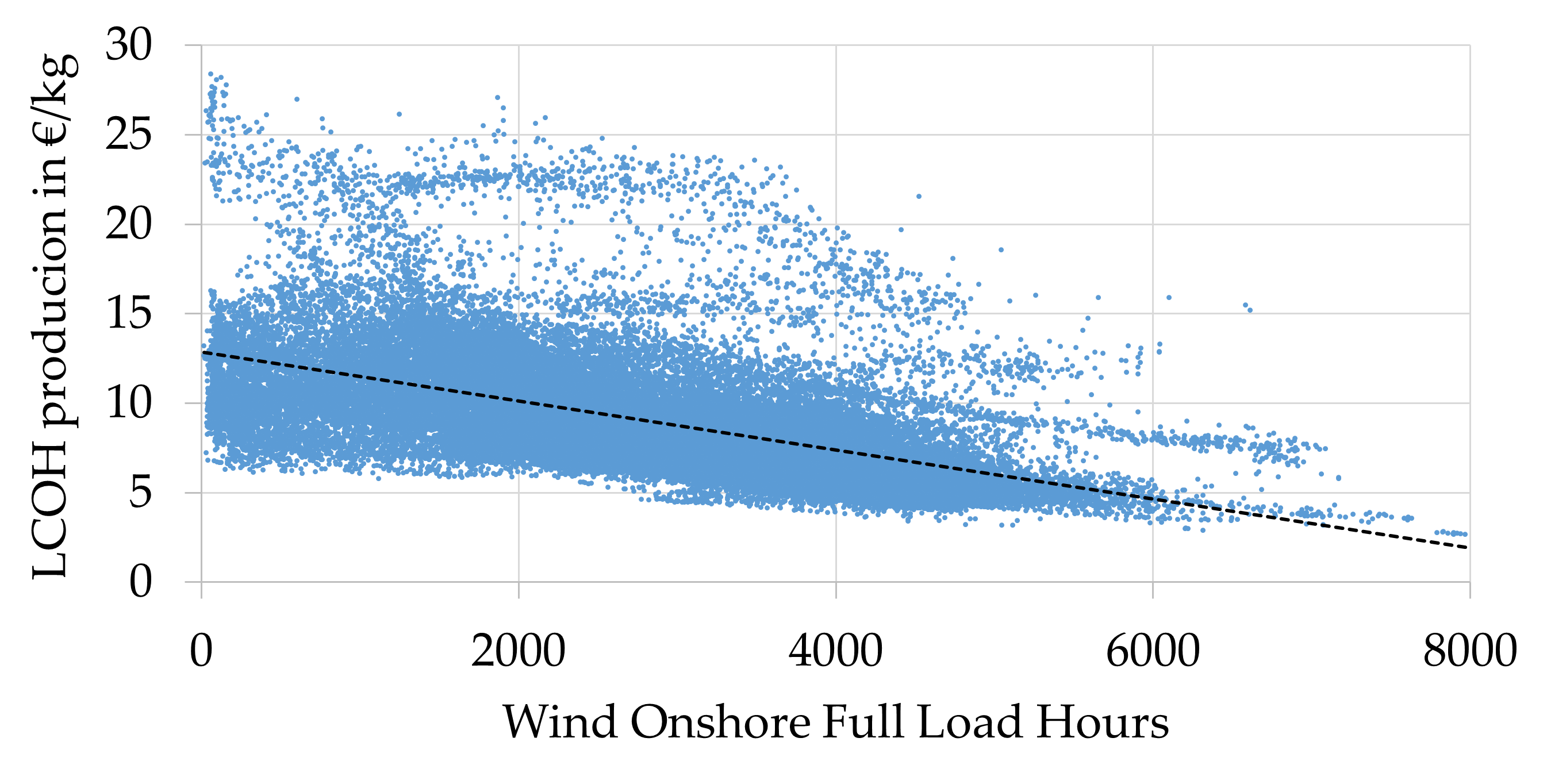} 
    \caption{FLH vs. LCOH production for wind onshore in the scenario BASE} \label{fig:FLH cWACC and BASEd}
  \end{subfigure}

\caption{Correlation between FLH and LCOH production for the two renewable energy technologies wind onshore and PV in the scenarios cWACC and BASE.}
\label{fig:FLH cWACC and BASE}
\end{figure}

\begin{figure}
  \begin{subfigure}{0.49\textwidth}
    \includegraphics[width=\linewidth]{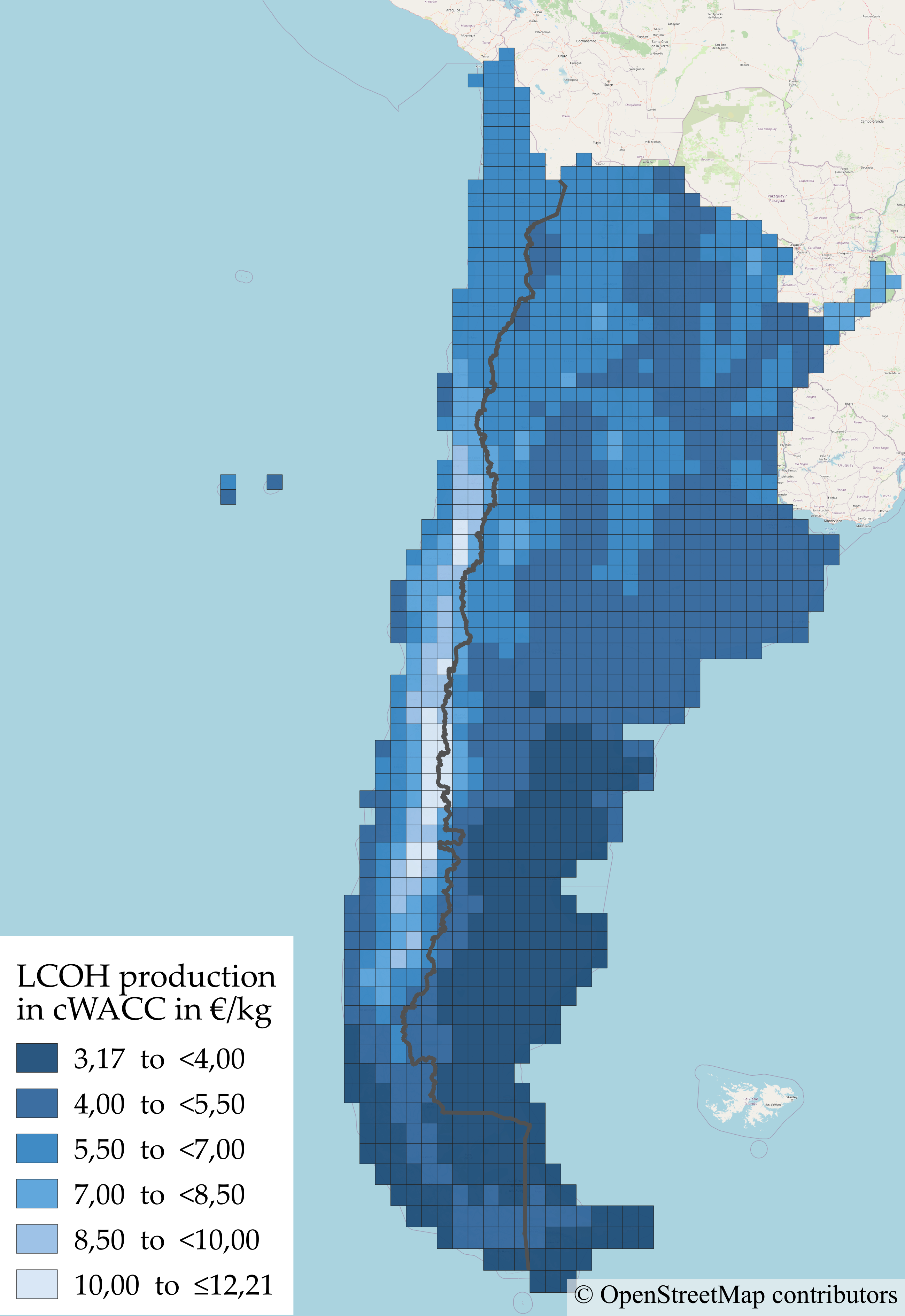} 
    \caption{Scenario cWACC} \label{fig:LCOH Chile Argentina_a}
  \end{subfigure}
  \hspace*{\fill}   
  \begin{subfigure}{0.49\textwidth}
    \includegraphics[width=\linewidth]{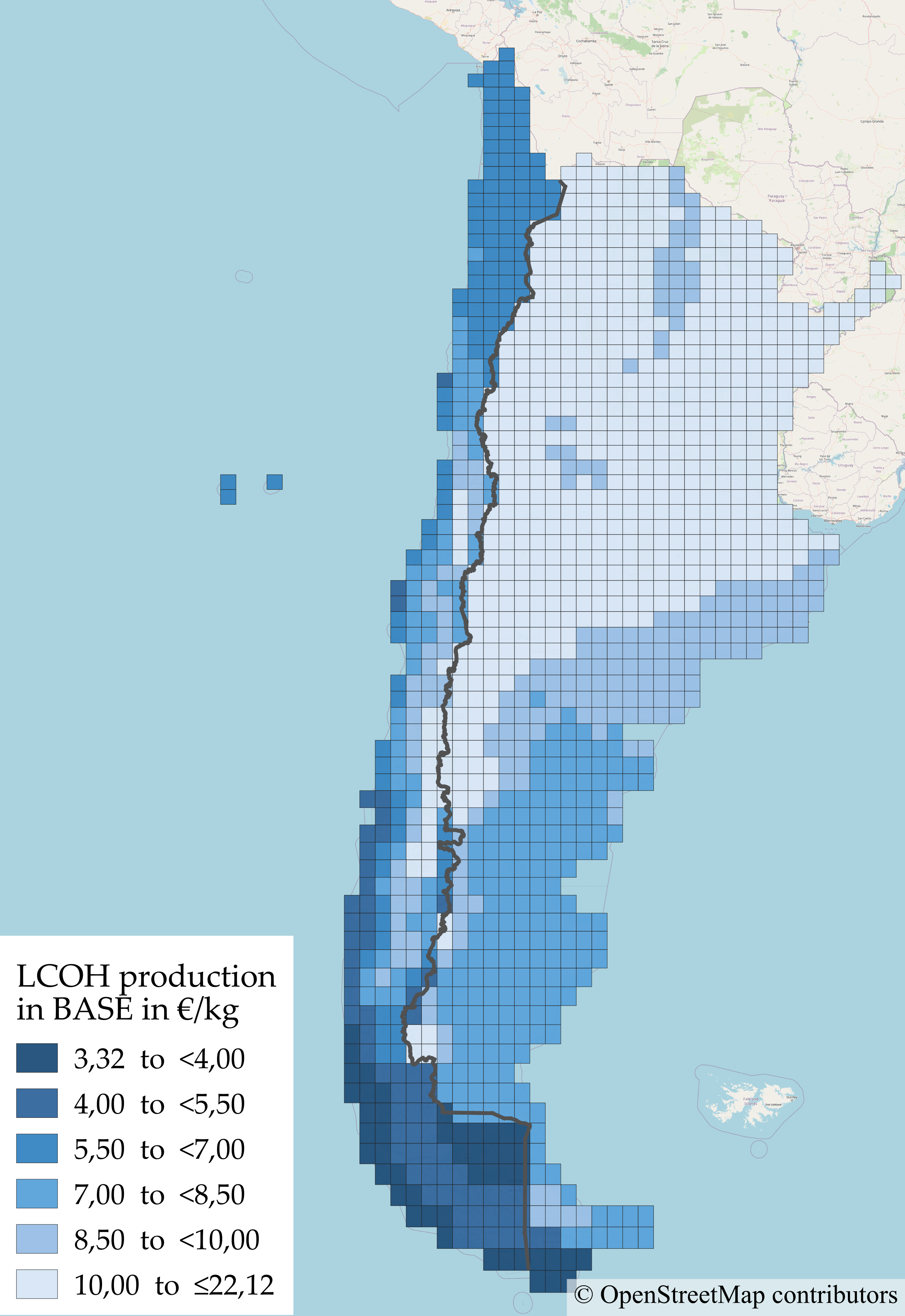} 
    \caption{Scenario BASE} \label{fig:LCOH Chile Argentina_b}
  \end{subfigure}

\caption{LCOH production in Chile and Argentina on a \qtyproduct{50x50}{\km} grid for the scenarios cWACC and BASE.}
\label{fig:LCOH Chile Argentina}
\end{figure}

Figure~\ref{fig:cWACC and BASEb} shows the LCOH production in the scenario BASE. When comparing Figure~\ref{fig:cWACC and BASEa} to Figure~\ref{fig:cWACC and BASEb}, it becomes evident that the CRPs influence the LCOH production besides the meteorological conditions. In some countries, such as Argentina, Brazil, or countries in middle and northern Africa, the influence of the CRPs counteracts good renewable energy resources with high FLH, i.e., even though the FLH are high the large CRPs lead to uncompetitive LCOH production. This can also be seen in Figure~\ref{fig:FLH cWACC and BASEc} and \ref{fig:FLH cWACC and BASEd}. The correlation between FLH and low LCOH production becomes less evident in the scenario BASE than in cWACC. Even though the correlation is still significant at a 1\% level, correlation factors for wind onshore and PV decrease to -0.58 and -0.06, respectively. Linear regression still shows a negative relation on a 1\% significance level, but R-squared values are lower (see Table~\ref{tab:LinRegTable}), meaning that FLH explain less of the variation in LCOH production compared to the scenario cWACC. The most compelling example involves Chile and Argentina, with similar renewable energy resources but different economic situations: Chile's CRP of 0.68\% is well below Argentina's CRP of 11.62\%~\cite{CRP.2021}. Figure~\ref{fig:LCOH Chile Argentina} illustrates the effect of the CRP on the LCOH production for both countries in detail. Between the two scenarios cWACC (Figure~\ref{fig:LCOH Chile Argentina_a}) and BASE (Figure~\ref{fig:LCOH Chile Argentina_b}), a clear difference regarding the LCOH production can be perceived in Argentina. When considering CRPs, the LCOH production increases by 101\% to 117\%, meaning that the production cost for green hydrogen more than doubles. The cheapest LCOH production increases from 3.5~\geneuro/kg to 7.3~\geneuro/kg. In Chile, on the contrary, the LCOH production does not significantly rise when considering CRPs. Globally, LCOH production increases due to the impact of the CRPs. As in the scenario cWACC, the lowest LCOH production can be found in the Kerguelen Islands in the Indian Ocean (2.7~\geneuro/kg). In contrast, the highest LCOH production can now be found in Venezuela (28.4~\geneuro/kg). While the CRP only has a minor effect on the LCOH production in the cheapest cell, as the Kerguelen Islands belong to France, where the CRP is low, the high CRP of Venezuela (19.18\%) superimposes the relatively moderate LCOH production in Venezuela in the scenario cWACC (max. LCOH production of 9.9~\geneuro/kg). Therefore, the cell with the highest LCOH production changes from Canada to Venezuela. In the scenario BASE, the mean LCOH production rises to 9.1~\geneuro/kg, and the median is 8.6~\geneuro/kg. The minimum, maximum, average, and median LCOH production for all four scenarios can be found in Table~\ref{tab:Scenarios Min Max Average}.

\begin{table}[]
    \footnotesize
    \centering
    \caption{Minimum, maximum, median, and average LCOH production in cWACC, BASE, PVonly, and WINDonly}
    \label{tab:Scenarios Min Max Average}
    \begin{tabular}{p{1.8cm} p{1.7cm} p{1.7cm} p{1.7cm} p{2cm}}
        \toprule
        \multicolumn{1}{m{1.8cm}}{} & \multicolumn{1}{m{1.7cm}}{cWACC \newline in \geneuro/kg} & \multicolumn{1}{m{1.7cm}}{BASE in \geneuro/kg} & \multicolumn{1}{m{1.7cm}}{PVonly in \geneuro/kg} & \multicolumn{1}{m{2cm}}{WINDonly in \geneuro/kg} \\
        \midrule
        Minimum & 2.6 & 2.7 & 5.8 & 2.7 \\

        Median & 6.9 & 8.6 & 12.8 & 11.9 \\
        
        Average & 7.2 & 9.1 & 15.4 & 21.2 \\

        Maximum & 25.7 & 28.4 & 55.6 & 1592.5 \\
        \bottomrule
    \end{tabular}
\end{table}

Even though \cite{EWI, PtXAtlas, IRENA.2022c, AFRY.2022} calculate the LCOH production for future years and therefore differing cost assumptions, the geographic location of most of the more favorable hydrogen production sites coincides with the identified regions in the scenario BASE. Some of these countries are Chile~\cite{EWI, PtXAtlas, IRENA.2022c, AFRY.2022}, the USA~\cite{EWI, IRENA.2022c}, Australia~\cite{IRENA.2022c, AFRY.2022} Saudi Arabia~\cite{IRENA.2022c, AFRY.2022} and China~\cite{EWI, IRENA.2022c}. Also, Canada~\cite{Agora.2021}, Mauritania~\cite{PtXAtlas} and Northern Europe~\cite{EWI} represent favorable regions for low LCOH production in different studies and the scenario BASE. Given the high spatial resolution of the electrolysis-based HPSM, other locations, such as New Zealand, South Africa, Brazil, or Kenya, are additionally identified as locations with low LCOH production, not yet identified in the literature. 

While the CRPs impact the absolute level of the LCOH production, they do not influence the composition of the HPS in most cells. In more than 80\%, the share between installed wind and PV capacity remains within an absolute deviation of \textpm 1\%. In the remaining cells, the percentage of installed PV capacity changes from -27\% to 30\%. Since the CRP's effect on the LCOH production is not linear due to the annualized costs (see Equation~\ref{equ:Annual Cost}), its effect on the HPS design is higher in cells where high CRPs apply. While in countries with low CRPs the lifetime of the components barely influences the annualized costs, in countries with high CRP the components with higher lifetime will have higher annualized costs relative to components with smaller lifetimes. The PEM electrolyzer and compressor can be identified as main drivers for the changing composition as their lifetimes are shorter than those of the renewable electricity production technologies, and, therefore, their costs are less affected by the higher WACC. Similarly, hydrogen storage has a longer lifetime than renewable generation technologies and is therefore more affected by the higher WACC, which affects the HPS's composition.

\begin{figure}[h]
    \centering
    \includegraphics[width=0.9\textwidth]{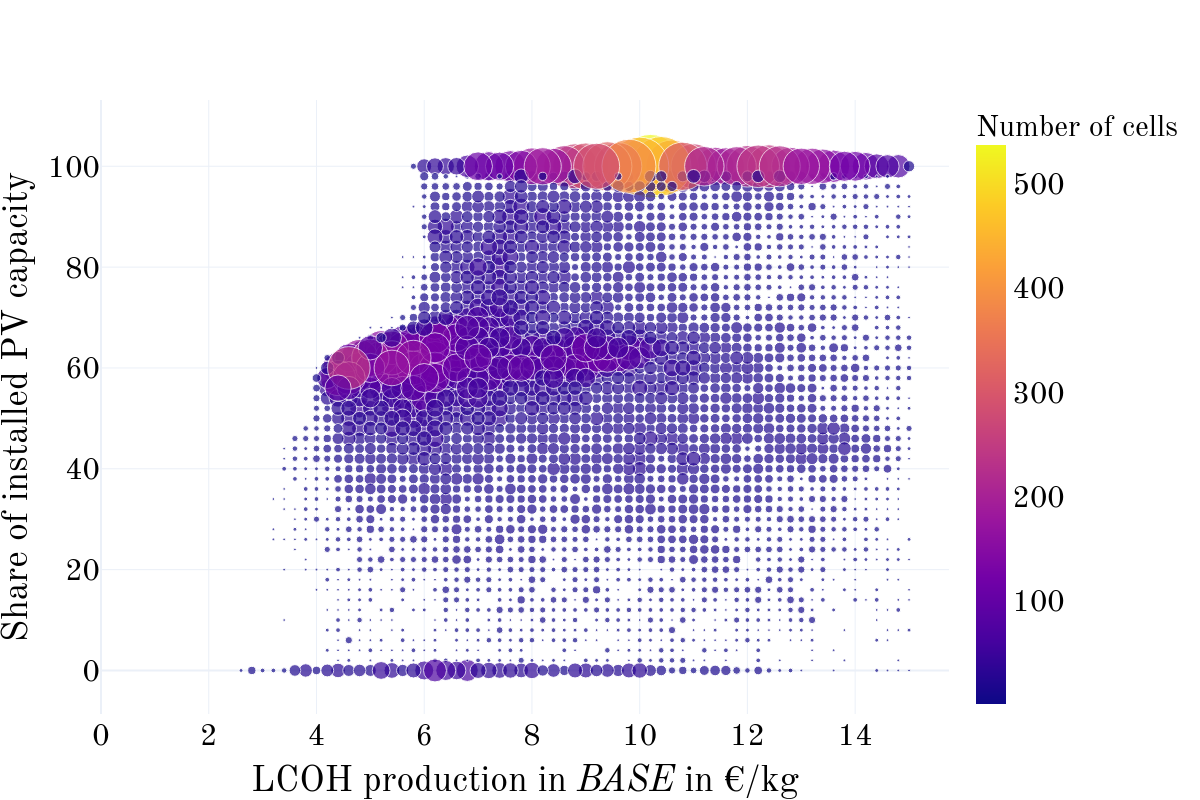} 
    \caption{Share of installed PV capacity in regard to the total installed renewable energy capacity for cells with LCOH production below 15~\geneuro/kg in the scenario BASE. Discretization is realized with an interval of 0.2~\geneuro/kg and 2\% PV share.}
    \label{fig:PVshare}
\end{figure}

\begin{figure}
  \begin{subfigure}{0.49\textwidth}
    \includegraphics[width=\linewidth]{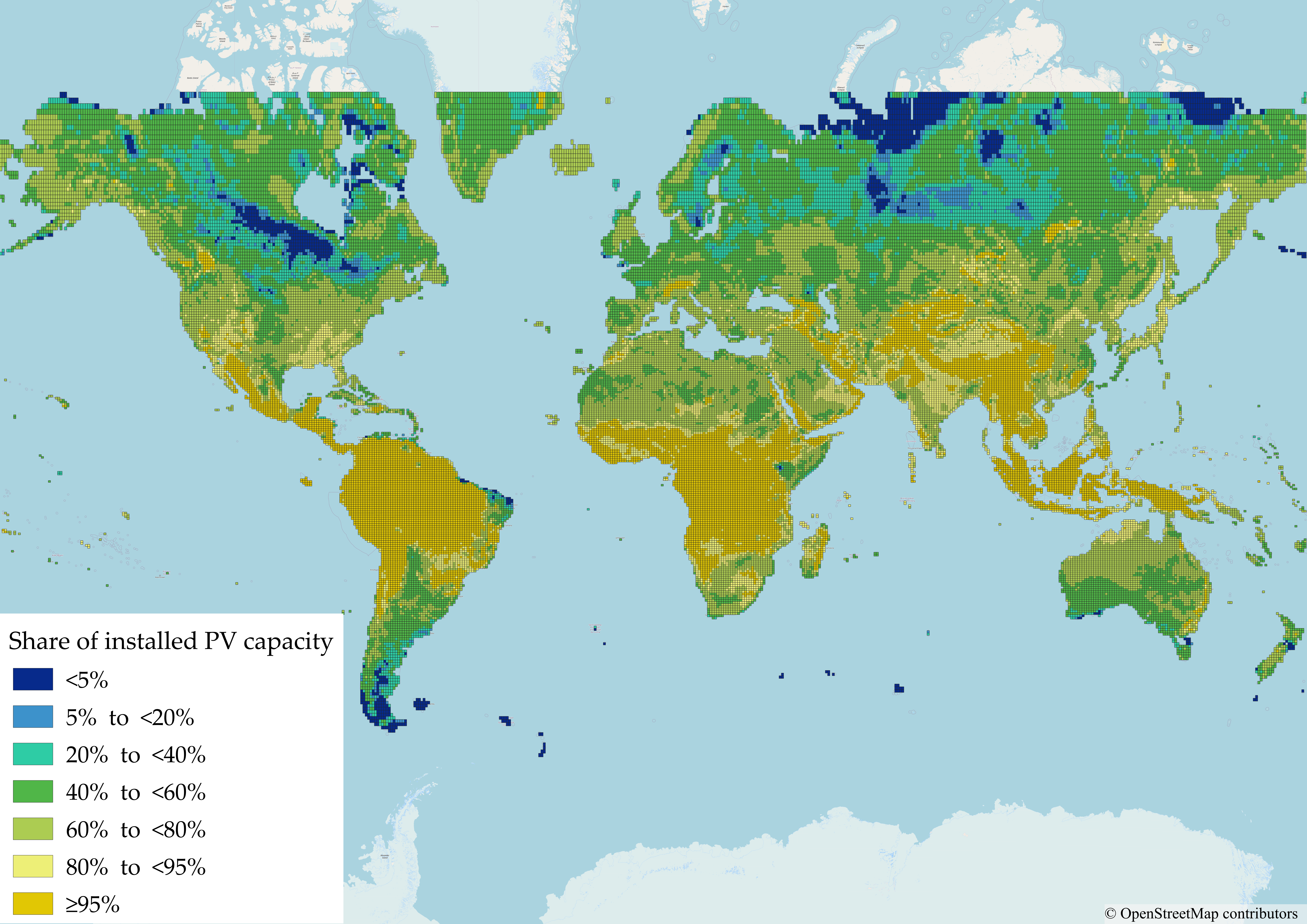} 
    \caption{Share of installed PV capacity in regard to the installed total renewable energy capacity per cell.}
    \label{fig:PVshare Costs a}
  \end{subfigure}
  \hspace*{\fill}   
  \hfill
  \begin{subfigure}{0.49\textwidth}
    \centering
    \includegraphics[width=0.75\textwidth]{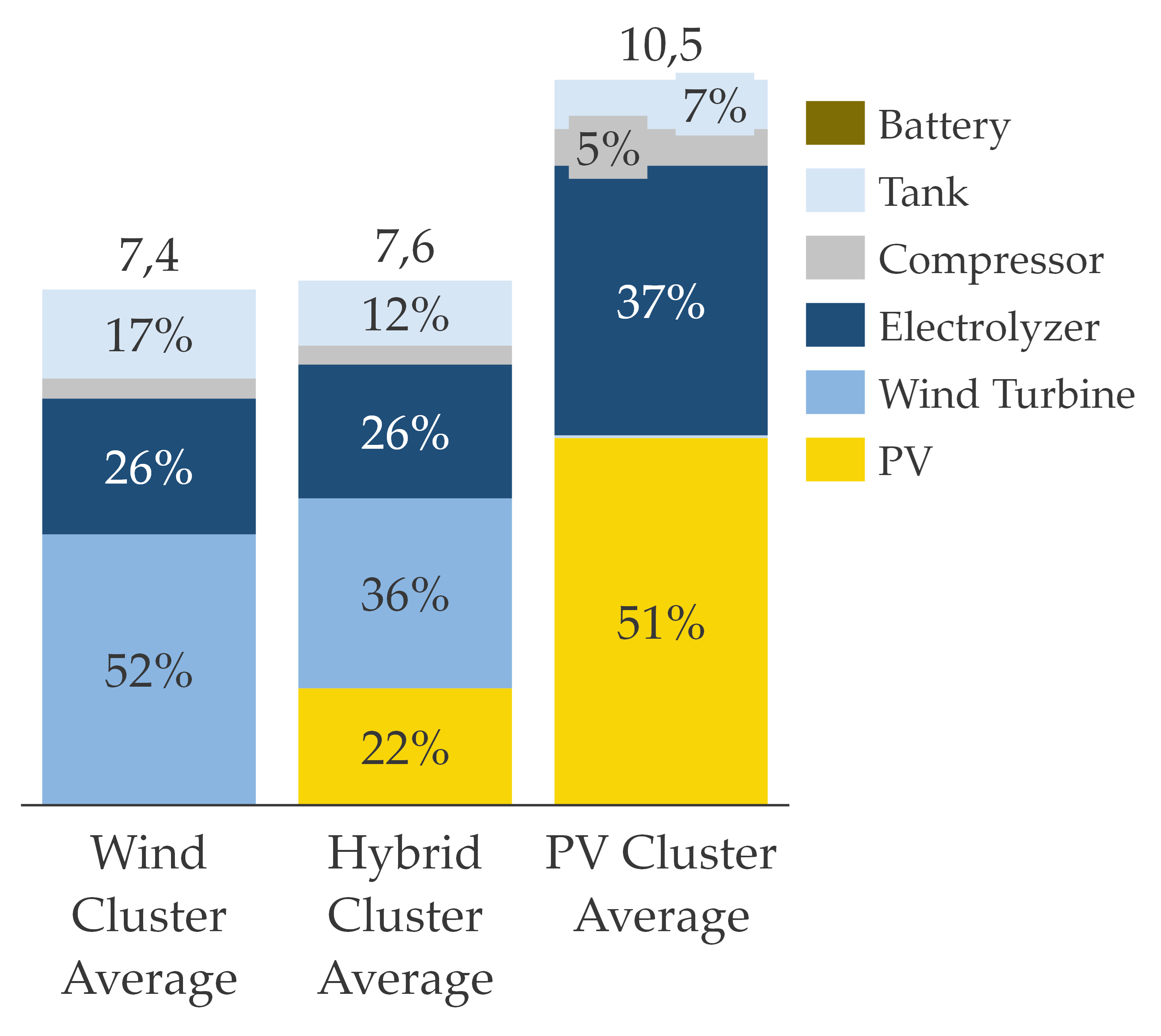}  
    \caption{Contribution of the individual HSP's components to the total LCOH production per cluster.}
    \label{fig:PVshare Costs b}
  \end{subfigure}

\caption{Composition of the hydrogen production systems in the scenario BASE.}
\label{fig:PVshare Costs}
\end{figure}

In Figure~\ref{fig:PVshare}, the share of installed PV capacity in regard to the total renewable energies capacity installed is plotted against the LCOH production for each cell, as well as the absolute number of cells with the same share of installed PV capacity. LCOH production below 4~\geneuro/kg can only be found in 173~cells~(0.33\%). These are cells with wind onshore as the primary or only renewable energy resource. Here, the hybrid lowest-cost constellations can be found in Australia and Greenland. To better analyze the components' contribution to the total LCOH production, we define three clusters focusing on LCOH production below 15~\geneuro/kg as more than 95~\% of the cells in the scenario BASE lay below this value and low LCOH production is a prerequisite for future development. The wind cluster with a share of installed PV capacity regarding the total installed renewable energy's capacity below 5\% (Wind Cluster), the hybrid cluster with a share of installed PV capacity in regard to the total installed capacity between 50\% and 70\% (Hybrid Cluster), and the PV cluster with a share of installed PV capacity in regard to the total installed capacity above 95\% (PV Cluster). About two thirds of all cells lay within these clusters: 2.5\% in Wind Cluster, 40.0\% in Hybrid Cluster, and 19.4\% in PV Cluster. Cells within the PV Cluster are situated in regions with good PV but bad wind conditions. Most are located near the equator (see Figure~\ref{fig:PVshare Costs a}).

 The average LCOH production in the Wind Cluster is 7.4~\geneuro/kg, 7.6~\geneuro/kg in the Hybrid Cluster, and  10.5~\geneuro/kg in the PV Cluster in the scenario BASE. The effect of the CRPs on the HPS design in the three clusters compared to the scenario cWACC is small, as only 13.1\% of the cells in these three clusters see high deviations in their installed capacities. Interestingly cells in the PV Cluster with high shares of installed PV have significantly higher CRPs than cells in the Wind Cluster. This  explains why the effects of CRPs on the correlation between FLH and LCOH production in Figure~\ref{fig:FLH cWACC and BASE} are stronger for PV than wind onshore. In summary CRPs mainly influence the LCOH production and not the optimal HPS design.   

\begin{figure}[h]
    \centering
    \includegraphics[width=1\textwidth]{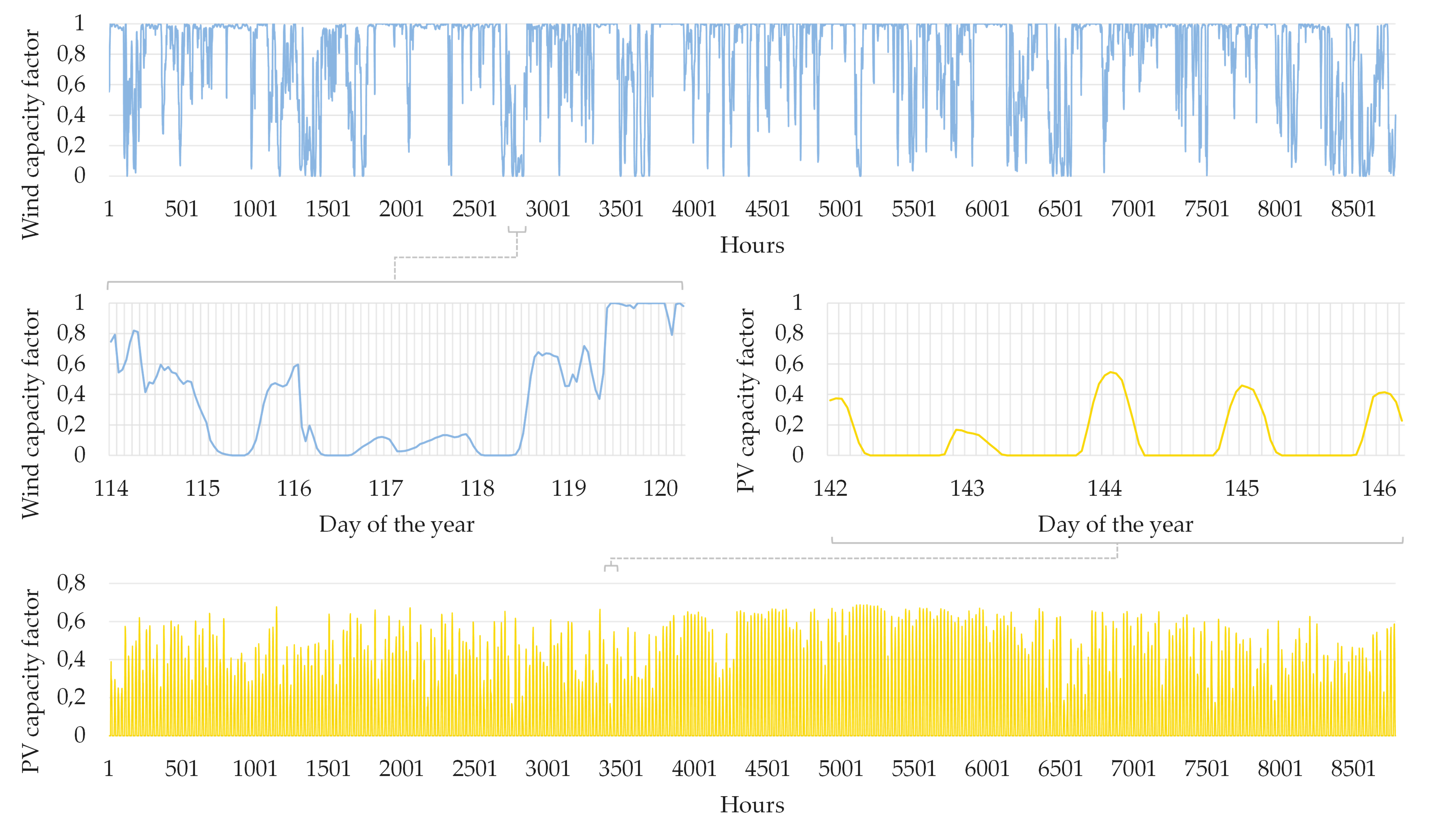} 
    \caption{Hourly capacity factor time-series for a representative cell in the Wind Cluster in Chile and a cell in the PV Cluster in Brazil.}
    \label{fig:CFTimeseries}
\end{figure}

 Figure~\ref{fig:PVshare Costs b} shows the mean contribution of the individual HSP's components regarding the total LCOH production for each of the three clusters. As can be seen, the costs for renewable electricity production account for more than 50\% of the total LCOH production in all clusters. The second largest cost component in all clusters is attributable to the electrolyzer. It is striking that the cost contributions of the electrolyzer in the Wind Cluster and Hybrid Cluster are almost identical but account for a significantly higher proportion of the already higher total LCOH production in the PV Cluster. On the other hand, the cost share of the hydrogen tank decreases with an increase in installed PV capacity while that of the compressors increases. This can be explained by the fact that in the Wind Cluster, larger hydrogen storage tanks are necessary to bridge periods with low winds and, therefore, small electricity production. In the PV Cluster, however, hydrogen storage volumes are lower. Thus, the costs for hydrogen storage do not contribute as strongly as in the Wind Cluster. However, storage is used more frequently, as production downtimes during the night must be compensated for. This is also why the dimensioning of the electrolyzer turns out to be larger. More hydrogen must be produced in less time to bridge supply and demand during the night. Figure~\ref{fig:CFTimeseries} shows the capacity factors time series for a representative cell in the Wind Cluster in Chile and a representative cell in the PV Cluster in Brazil. While the cell in the Wind Cluster presents periods of low wind, the cell in the PV Cluster does not have long periods with low capacity factors. In all three clusters and the remaining cells, battery storages are being deployed in very few cells (<0.1\%) and play no role in the HPS design, as they are too expensive compared to other component combinations. This result is in line with the findings of \cite{Fasih}. At a majority of cells the weak wind turbines for wind onshore are preferred over the strong wind turbines. 

\subsection{Scenarios PVonly and WINDonly}

As could be seen in the scenario BASE, in the majority of cells, hybrid HPSs are the most cost-optimal solution. To evaluate the cost reduction of hybrid HPSs, we calculated the PVonly and WINDonly scenarios, where only wind onshore or PV are available to the HPSM as a renewable electricity production technology. The LCOH production increases in both scenarios compared to BASE (see Table~\ref{tab:Scenarios Min Max Average}). While the minimum LCOH production in WINDonly remains the same as in BASE, the minimum LCOH production in PVonly more than doubles to 5.8~\geneuro/kg. This can be explained by the fact that the cell with minimum LCOH production in the scenario BASE corresponds to a cell dominated by wind onshore as the electricity production technology. On the contrary, the average LCOH production shows a stronger increase compared to the scenario BASE, with 6.3~\geneuro/kg and 12.1~\geneuro/kg in PVonly and WINDonly, respectively. 

\begin{figure}[h]
    \centering
    \includegraphics[width=0.8\textwidth]{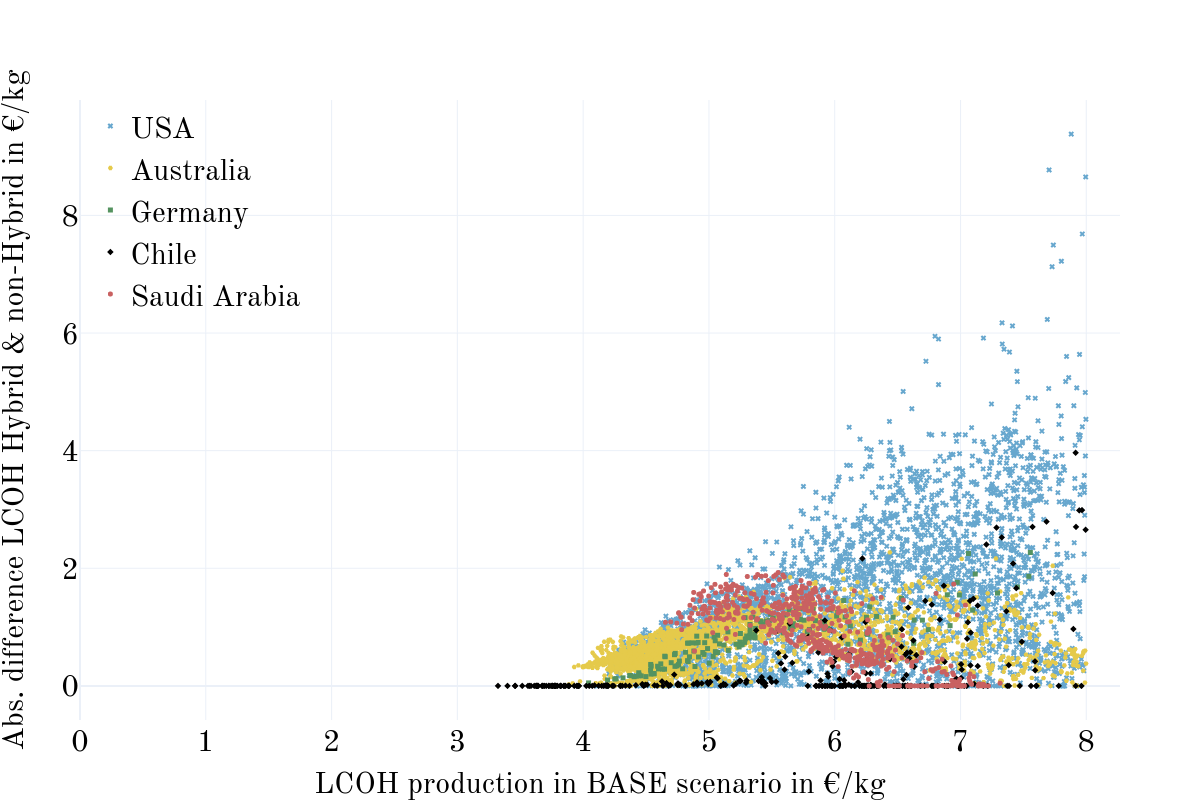}
    \caption{Difference in the LCOH production between hybrid (BASE) and non-hybrid design (PVonly and WINDonly) for Australia, Chile, Germany, Saudi Arabia, and the USA.}
    \label{fig:Diff Hybrid PVWIND}
\end{figure}

In the following, we analyze the LCOH production differences between hybrid designs from the BASE scenario and the non-hybrid design from PVonly and WINDonly. The minimum LCOH production for each cell in the non-hybrid scenarios PVonly and WINDonly is defined as the minimum LCOH production of each cell in both scenarios. Figure~\ref{fig:Diff Hybrid PVWIND} shows the absolute difference in the LCOH production for Australia, Chile, Germany, Saudi Arabia, and the USA. From these, Australia, Saudi Arabia, and the USA are the countries where hybrid HPSs significantly impact LCOH production.

Cells with low LCOH production show small cost difference between hybrid and non-hybrid designs. This effect can be explained by the fact that cheaper cells are dominated by one technology, most of the time wind, already. Chile is the most pronounced example, as most of Chile's cells with low LCOH production are dominated by wind generation. While the cost reduction in Australia, Chile, Germany, and Saudi Arabia is limited to 2~\geneuro/kg in most cases (99.5\% of all cells within the countries), hybrid HPSs can reduce the LCOH production by more than 2~\geneuro/kg in 25.4\% of the cells in the USA. While a hybrid HPS design does not significantly reduce LCOH production in cells with good renewable energy resources, it can significantly lower LCOH production in cells with less favorable renewable energy conditions, enabling competitive LCOH production in more regions of the world. 

\section{Discussion}
\label{sec:Discussion}
Even though the presented approach yields valuable insights into global LCOH production, it has some limitations. The presented method cannot replace site-specific planning when implementing HPSs. While the chosen spatial resolution of \qtyproduct{50x50}{\km} is high for a global analysis, it is coarse from a meteorological point of view when considering one specific site within one cell. Wind and PV parks rarely exceed the size of several square kilometers and are, therefore, smaller than the model's spatial resolution. Hence, meteorological conditions may be more or less favorable at specific sites within one cell, possibly affecting the LCOH production. Additionally to weather data, the used CRP data has a strong effect on the LCOH production, as shown in Section~\ref{sec:Results}. The used CRPs and WACC are not asset or project-specific and can only give an indication on the country and project-specific cost of financing. Additionally state guarantees or similar measure can have impact on project specific cost of financing~\cite{IRENA.2023,Moritz.2022}. Survey based data from \cite{IRENA.2023} for selected countries provides cost of financing in a similar order of magnitude for wind onshore and PV technologies. In \cite{Moritz.2022} methodology from \cite{EWI} is extended to consider WACC derived from oil and gas industry. The used source data could not be accessed for comparison.

Nevertheless, the presented approach provides reasonable estimates for favorable HPS locations on a global scale. The estimate would even be better if the LCOH production were calculated with different weather years, even though cell size already smoothes weather year-dependent effects. Due to the computational expenses, this was not able in this paper yet. Land and water availability are a further restriction not yet considered in the model. Natural reserves, densely populated regions, or geological conditions may inhibit the construction of a HPS within the \qtyproduct{50x50}{\km} cell. As the meteorological data limits the model's resolution, land availability will not affect the LCOH production but the total amount of hydrogen possibly produced within one cell. Therefore, no statement can be made regarding the hydrogen production potential within one cell. The constant demand of 1~kg of hydrogen per hour influences the LCOH production, e.g., the model  curtails electricity generated but not needed to cover the continuous demand of 1~kg/h. Pre-analysis shows that with existing HPS sizes, up to 34\% of additional hydrogen could be produced with curtailed electricity, leading to a LCOH production reduction. Furthermore, allowing the HPS to temporarily not satisfy demand could lead to smaller needed system components and therefore reduce LCOH production. Further research on hydrogen demand flexibility is needed to quantify these effects. 

As the model does not consider a connection to the public grid (excess) electricity cannot be sold or bought to and from the grid, which could lead to reduced LCOH production. Regarding the HPS itself, costs for planning and construction and the degradation of the stack have been neglected. Both would increase the total LCOH production, as the cost of planning and construction must be added to the total system costs. At the same time, the degradation of the stack reduces the electrolyzer's efficiency and, therefore, the hydrogen output over the lifetime of the HPS.  

To estimate the total global potential for renewable hydrogen production, offshore hydrogen production sites should also be considered. However, the analysis is complex, as data regarding the bathymetry is needed, and construction limitations due to ocean depth must be taken into account. As additional cells are identified, the computational demand increases, too. Nevertheless, the model adaptions discussed must be considered to better understand hydrogen production and how it will interact with the steeply rising demand in the future to aid in decarbonizing global energy demand.  

\section{Conclusion}
\label{sec:Conclusion}
We present a method to systematically model LCOH production with high spatial resolution on a global level considering hybrid HPSs and country-specific interest rates. LCOH production and optimal HPS designs are calculated for more than 50,000 inland locations using a cost optimization approach and hourly weather data on a \qtyproduct{50x50}{\km} global grid. For the implementation, the open-source framework PyPSA~\cite{PyPSA} is used. Four scenarios are calculated to analyze the optimal HPS design and the influence of CRPs on the LCOH production.

CRPs show a strong effect on LCOH production. Our study confirms that already-known regions from the literature have low LCOH production. Furthermore, our study identifies additional regions with low LCOH production. Lowest-cost locations have low CRPs and high FLH, e.g., Chile. Considering CRPs at locations with high FLH but also high CRPs shows significant increases in LCOH production. This cost increase induced by CRPs overshadows low LCOH production due to good renewable energy resources in some countries, with Argentina being a prominent example. Locations with high shares of installed PV capacity show stronger effects of the CRP on LCOH production than locations with high shares of wind capacity. 

Findings show that hybrid HPSs consisting of both wind and PV as electricity production sources can significantly reduce the LCOH production in comparison to non-hybrid systems (Scenarios PVonly and WINDonly), enabling competitive LCOH production in more regions of the world. In the scenario BASE, more than one third (40.0\%) of all cells with a LCOH production below 15~\geneuro/kg have an installed PV capacity share between 50\% and 70\% of the total installed renewable energies capacity. CRPs showed little effect on the optimal HPS design. Hybrid HPSs lead to significant LCOH reductions for average and high LCOH production locations and insignificant cost reductions at locations with LCOH production below 4~\geneuro/kg that are dominated by one generation technology.

To compensate for fluctuations in hydrogen production due to the volatile renewable energy sources, hydrogen storage is deployed to balance production and  constant hourly demand. Battery storage, on the other hand, is not used, as their deployment is too expensive compared to alternative HPS designs.

In summary we show that CRPs and hybrid HPSs designs do have a significant impact on the LCOH production. Therefore, they must be considered to identify regions with low LCOH production.

\section*{Declaration of Competing Interests}
The authors declare that they have no known competing financial interests or personal relationships that could have appeared to influence the work reported in this paper.

\section*{Acknowledgement}
This work was funded by the German Federal Ministry for Economic Affairs and Climate Action (003EWT001A).

\appendix
\section{Appendix}
\label{sec:appendix}

\begin{table}[H]
    \footnotesize
    \centering
    \caption{Impact of FLH on LCOH production.}
    \label{tab:LinRegTable}
    \begin{tabular}{p{2cm} p{2cm} p{1.8cm} p{2cm} p{1.8cm} p{2cm}}

         \toprule
        & Scenario & \multicolumn{2}{c}{cWACC} & \multicolumn{2}{c}{BASE}\\
            \cmidrule{2-6}
        \multicolumn{1}{m{2cm}}{Variables} & \multicolumn{1}{m{2cm}}{Coef.} & \multicolumn{1}{m{2cm}}{Model PV} & \multicolumn{1}{m{2cm}}{Model Wind} & \multicolumn{1}{m{2cm}}{Model PV} & \multicolumn{1}{m{2cm}}{Model Wind}\\
         \midrule

        FLH PV & $\beta_{1}$ [PV] & -0.003*** (0.000) &  & -0.001*** (0.000) & \\
        
        FLH WIND & $\beta_{1}$ [Wind]  &  & -0.001*** (0.000) &  & -0.001*** (0.000)\\

        $R^{2}$ & -  & 14.25\% & 49.98\% & 0.36\% & 33.57\%\\

        adj. $R^{2}$ & - & 14.25\% & 49.98\% & 0.36\% & 33.57\%\\

        Obs. (N) & -  & 51677 & 51677 & 51677 & 51677\\

        \bottomrule
        \multicolumn{6}{p{14cm}}{\scriptsize Notes: The p-values are in parentheses and the significance levels are at the 1\%(***), 5\%(**), and 10\%(*)}\\  

        \multicolumn{6}{p{14cm}}{\scriptsize Estimated Model: $LCOH_{i} = \beta_{0} + \beta_{1} FLH_{i} + \epsilon{i}$, where $i$ is the number of calculated cells.}\\ 
    \end{tabular}
\end{table}

 \bibliographystyle{elsarticle-num} 
 \bibliography{cas-refs}





\end{document}